# The COVID-19 Pandemic and the Future of Telecommuting in the United States

**Deborah Salon**
School of Geographical Sciences and Urban Planning, Arizona State University, Tempe, AZ 85281
Email: dsalon@asu.edu

**Laura Mirtich**
School of Geographical Sciences and Urban Planning, Arizona State University, Tempe, AZ 85281
Email: lmirtich@asu.edu

**Matthew Wigginton Bhagat-Conway**
Department of City and Regional Planning, University of North Carolina at Chapel Hill, Chapel Hill, NC 27599
Email: mwbc@unc.edu

**Adam Costello**
School of Geographical Sciences and Urban Planning, Arizona State University, Tempe, AZ 85281
Email: astiffe@asu.edu

**Ehsan Rahimi**
Department of Civil, Materials, and Environmental Engineering, University of Illinois at Chicago, Chicago, IL 60607
Email: erahim4@uic.edu

**Abolfazl (Kouros) Mohammadian**
Department of Civil, Materials, and Environmental Engineering, University of Illinois at Chicago, Chicago, IL 60607
Email: kouros@uic.edu

**Rishabh Singh Chauhan**
Department of Civil, Materials, and Environmental Engineering, University of Illinois at Chicago, Chicago, IL 60607
Email: rchauh6@uic.edu

**Sybil Derrible**
Department of Civil, Materials, and Environmental Engineering, University of Illinois at Chicago, Chicago, IL 60607
Email: derrible@uic.edu

**Denise da Silva Baker**
School of Sustainable Engineering and the Built Environment, Arizona State University, Tempe, AZ 85281
Email: denise.silva@asu.edu

**Ram M. Pendyala**
School of Sustainable Engineering and the Built Environment, Arizona State University, Tempe, AZ 85281
Email: Ram.Pendyala@asu.edu

**ABSTRACT**

This study focuses on an important transport-related long-term effect of the COVID-19 pandemic in the United States: an increase in telecommuting. Analyzing a nationally representative panel survey of adults, we find that 40-50% of workers expect to telecommute at least a few times per month post-pandemic, up from 24% pre-COVID. If given the option, 90-95% of those who first telecommuted during the pandemic plan to continue the practice regularly. We also find that new telecommuters are demographically similar to pre-COVID telecommuters. Both pre- and post-COVID, higher educational attainment and income, together with certain job categories, largely determine whether workers have the option to telecommute. Despite growth in telecommuting, approximately half of workers expect to remain unable to telecommute and between 2/3 and 3/4 of workers expect their post-pandemic telecommuting patterns to be unchanged from their pre-COVID patterns. This limits the contribution telecommuting can make to reducing peak hour transport demand.

**MOTIVATION**

The COVID-19 pandemic gave many workers extended experience with working from home. They learned how to telecommute effectively, and they learned what they do and do not like about it. Employers also learned how to manage remote employees, and they learned what they do and do not like about having remote employees. Because both workers and employers learned that they like some aspects of remote work arrangements, many observers have argued that there will be a permanent shift toward more telecommuting, even after the pandemic is behind us.

When US workers can safely return to the workplace, will they return to their pre-COVID commuting patterns? If not, who among them will adopt a "new normal"? How might this change the environmental impact of our transportation system? What are the implications for worker productivity and quality of life? Will the "new normal" improve or diminish societal equity? These questions motivate this research.

Research on the determinants of telecommuting prior to the COVID-19 pandemic investigated the importance of employment characteristics (e.g., job type, industry, employer size, length of tenure, and employer policies), demographics (e.g., education, income, gender, household composition, and race/ethnicity), attitudes, and the built environment (e.g., accessibility, urban/rural) (Asgari et al., 2014; Singh et al., 2013; Tang et al., 2011). Telecommuters tended to be older (Peters et al., 2004; Thériault et al., 2005; Walls et al., 2007), male (Sener and Bhat, 2011; Thériault et al., 2005; Van Sell and Jacobs, 1994), and well-educated (Peters et al., 2004; Van Sell and Jacobs, 1994; Walls et al., 2007). A number of studies point out that the factors associated with having the option to telecommute may differ from those determining how often a worker actually chooses to do so (Pouri and Bhat, 2003; Sener and Bhat, 2011; Singh et al., 2013; Walls et al., 2007).

**Telecommuting's impact on travel**

Telecommuting has been widely studied as a strategy for reducing travel and its undesirable consequences such as traffic congestion and poor air quality. Despite hopes that remote work may be an effective travel reduction strategy, prior literature reports mixed findings. Some find that telecommuters travel about the same amount or even *more* than their counterparts who work in person (Allen et al., 2015; Choo et al., 2002; Zhu, 2012; Zhu et al., 2018). Others have found a reduction in travel associated with telecommuting (Choo et al., 2005; Henderson and Mokhtarian, 1996; Shabanpour et al., 2018; Zupan, 1994). These effects are modest, however, and some note that the reduction in commute miles for a small number of people is simply not a large enough change to produce substantial travel reductions across the population (Salon et al., 2012; Walls et al., 2005). An oft-cited reason for this underwhelming reduction in travel is that telecommuters, freed from a daily commute, are willing to live further from the workplace and therefore take longer trips when they do visit their workplaces or other destinations (Rhee, 2008). Even if the reduction in vehicle miles traveled is small from telecommuting, some authors emphasize that remote work still eases congestion by giving remote workers the freedom to drive at non-peak hours and can reduce the number of high-emission "cold starts" made by drivers (Shabanpour et al., 2018; Su et al., 2021; Zupan, 1994), thereby providing benefits even in the absence of substantial travel reduction.

**Telecommuting's impact on personal and professional well-being**

Aside from reducing travel, there is substantial interest in telecommuting as a means of improving individual outcomes for employees, both at work and in their personal lives. Numerous studies have found that pre-pandemic telecommuting was associated with higher worker productivity. Two literature reviews of the subject found that productivity gains are associated with telecommuting in most research (Allen et al., 2015; Van Sell and Jacobs, 1994). One study estimates productivity gains from telecommuting around 20% (Frolick et al., 1993), while others report a productivity increase of unspecified magnitude (Soenanto et al., 2016; Tustin, 2014). Especially before the COVID-19 pandemic, however, telecommuters were likely a self-selected group with personal and job characteristics exceptionally well-suited to remote work. This means that these pre-pandemic estimates of productivity gains may not apply to pandemic-era telecommuters.

Outside of the workplace, telecommuting also influences workers' personal lives. This influence is often positive, with remote workers reporting higher job satisfaction (Allen et al., 2015; Masayuki, 2018; Tustin, 2014; Van Sell and Jacobs, 1994), lower job turnover (Allen et al., 2015), and higher life satisfaction (Hornung and Glaser, 2009; Masayuki, 2018; Tustin, 2014) than in-person workers. However, some point out that fully remote work does not provide the best quality of life for many employees (Virick et al., 2010), and a small literature on work-family conflicts among telecommuters finds inconsistent patterns on whether such conflicts are eased or intensified by remote work (Allen et al., 2015; Hornung and Glaser, 2009; Sarbu, 2018). Moreover, some negative effects of telecommuting on well-being have been reported, including poor work-life balance (Grant et al., 2013; Rhee, 2008), social isolation (Allen et al., 2015; Bentley et al., 2016; Hager, 2018; Van Sell and Jacobs, 1994), and stress due to technological challenges (Tustin, 2014).

There have been a number of studies of telecommuting during the COVID-19 pandemic. Many have focused on disparities in access to telecommuting, since working from home during the pandemic protected workers from virus exposure. Key disparities in the ability to work from home during the pandemic included those between educational attainment groups, income groups, race/ethnicity groups, and job types (Bick et al., 2021; Gould and Kandra, 2021; Ray and Ong, 2020; Ward and Kilburn, 2020). Ray and Ong (2020) find that most of the race/ethnicity disparity can be explained by disparities in income, educational attainment, and job type. Barrero, Bloom, and Davis (2021) focus on the future of telecommuting, and argue that pandemic-era telecommuting will "stick" due to a combination of technological advances and investments that make telecommuting more effective and changes in people's attitudes toward telecommuting.

We contribute to the telecommuting literature with a study of how the COVID-19 pandemic has impacted both the option to telecommute and the choice of telecommuting frequency, both during the COVID-19 pandemic and looking to the post-COVID future. The main analysis and results reported here are based on Wave 1 of the COVID Future Panel Survey, a nationally representative survey dataset collected in the US by this research team between July and October 2020. As a check on the robustness of our results, we repeated all analyses presented here using Waves 2 and 3 of the COVID Future Panel Survey, which are smaller samples collected between November 2020 and April 2021, and in October and November 2021, respectively. None of the key results change, but where small changes are found, we identify them in the text. For full results tables for Wave 2 and Wave 3, please see the Appendix.

We reported in Salon et al. (2021b) that the fraction of US workers who expect to telecommute at least a few times per week is approximately double in the post-COVID period, compared to the pre-pandemic period. Here, we dive deeper to shed light on four related questions:

1. How did telecommuting patterns change during the pandemic and what do workers expect for the post-COVID future?

2. Among new telecommuters, for whom will pandemic-era telecommuting stick?

3. Who are the new telecommuters? How do they differ from both pre-COVID telecommuters and workers who never telecommuted?

4. Which factors are associated with telecommuting in the pre-COVID and pandemic eras, as well as expected telecommuting once COVID-19 is no longer a threat? Have these factors changed across these time periods?

Our research confirms findings from the existing literature regarding the factors associated with pre-COVID telecommuting and adds to this literature with analyses of telecommuting in the COVID era, worker expectations about telecommuting in the post-COVID future, and what these expectations mean for the future of transport sustainability, economic productivity, and quality of life.

**DATA**

The data that are the basis for this analysis come from the COVID Future Panel survey, collected using Institutional Review Board-approved survey instruments from a nationally representative sample of US adults between July and October of 2020 (Wave 1), between November 2020 and June 2021 (Wave 2), and in October and November 2021 (Wave 3). Respondents to Wave 1 were recruited using multiple methods, including a quota-sampled survey panel, direct random email invitations, media coverage of the research project, and additional volunteer participants who heard about the project from other participants. Only respondents to Wave 1 were invited to participate in Waves 2 and 3, and approximately one-third of the Wave 1 respondents responded to each subsequent Wave. The surveys included questions about a variety of topics related to the pandemic experience, including employment, remote work, worker productivity during the pandemic period, attitudes toward remote work, and detailed demographic information.

Telecommuting questions were asked three times in Wave 1: once about the pre-COVID period, once about the survey period, and once about expectations for the post-COVID future. In Waves 2 and 3, the questions were asked only about the survey period and the post-COVID future. For each period, respondents were first asked whether they had (or expected to have) the option to work from home in that period. Importantly, only those who answered yes were then asked how often they actually did or expected to work from home. These data therefore reflect what workers *want* to do tempered by what they are *able* to do based on the practical realities of their job and/or their employer's remote work policies. These data do not simply reflect stated preferences or wishful thinking about remote work. They are survey respondents' honest expectations about their own future.

In fact, Wave 3 included the question, "If your employer offered the option to work from home as much as you want after COVID-19 is no longer a threat, how much would you want to?" More than half of our respondents who reported that they expect not to be able to telecommute post-COVID also reported that they would prefer to telecommute sometimes (45%) or even every day (7%). Among respondents who expect to be able to telecommute, their preferred telecommuting frequency is largely the same as their expected telecommuting frequency. This offers further evidence that our survey data reflect our respondents' honest expectations about their own futures.

Because our focus here is on telecommuting, this analysis uses the portion of the survey sample that were employed before and/or during the pandemic (more than 4,500 respondents in Wave 1, and approximately 1,400 respondents in each of Waves 2 and 3), whether or not they worked from home. The data include weights that adjust summary statistic results to be representative of the US adult population in terms of age, education, gender, Hispanic status, household income, presence of children, number of household vehicles, and region of the US. For Waves 2 and 3, the included weights also adjust for nonresponse as well as additionally weighting for Wave 1 pre-pandemic telecommuting status. All summary statistics and marginal effects presented employ these weights. Multivariate statistical models control for most of these variables, and they were estimated without weights. Full details on the Wave 1 survey dataset and methodology are available in Chauhan et al. (2021a), and interested readers can download the dataset to conduct analyses of their own (Salon et al., 2021a). Other results based on these survey data can be found in (Chauhan et al., 2021b; Mirtich et al., 2021; Salon et al., 2021b; Silva et al., 2021). This article presents results based on the data from Wave 1B version 1.1, Wave 2 Main, and Wave 3 Main.

There are four limitations to this study that bear mentioning. First, the telecommuting frequency question for the during COVID period (i.e. survey period) was asked differently from this question in the pre- and post-COVID periods. Namely, in the during COVID period, respondents were asked to report how many days they worked from home in the last seven. For the pre- and post-COVID periods, they were asked to select among a set of "usual" frequency options for working from home. If a respondent telecommuted during the period in which they took the survey, but not in the seven days immediately prior, our survey mis-categorizes their current period telecommuting frequency.

Second, attitudes toward working from home were measured at the time of the survey, but are likely affected by pandemic era telecommuting experience. They may, therefore, be incorrectly measured for the pre-pandemic period model. Third, the survey was distributed only in English and only online, which may limit the representativeness of the sample for certain subpopulations.

Finally, we note that decisions about telecommuting are made jointly by workers and employers. Although we made efforts to ensure that our data incorporated worker expectations about future telecommuting informed by employer policies, it is a limitation of this work that we only collected data from workers and not from employers.

It is also important to keep in mind that the results reported here about the post-COVID future are based on people's expectations. It may be that unforeseen events will cause people's actual future choices to deviate from what they expected. Thus, future data collection will be needed to verify the post-COVID results of the COVID Future Panel Survey. That said, our survey respondents' expectations about whether they will be able to and how much they will choose to telecommute post-COVID have been stable or moving in the direction of expecting more telecommuting over three waves of data collection (see Appendix Table A-1). It would be surprising if telecommuting levels ever return to those seen in 2019.

**RESULTS**

**Telecommuting in three periods**

**Figure 1** provides an overview of the evolution of telecommuting ability and frequency in each of three periods based on the Wave 1 dataset: pre-COVID, during COVID, and post-COVID expectations. In each period, and throughout this article, the "Sometimes" telecommuting frequency category includes survey responses of "a few times/month," "once/week," and "a few times/week." The COVID Future survey language identified the post-COVID period as a future time when COVID-19 is "no longer a threat." The questions about expected telecommuting in the future were asked of survey respondents who were employed in either the pre-COVID period or at the time when they took the survey, and this is the sample used to generate this figure. For this reason, both the pre-COVID period and the during COVID period include an Unemployed category. The colors of the flows in **Figure 1** track the telecommuting category that respondents reported for the pre-COVID period.

The fraction of workers who are unable to telecommute is the largest group in every period, even during the COVID pandemic when including those who became unemployed. This is chiefly because many jobs require physical presence at the workplace. Our finding that between 50 and 65 percent of all US workers are unable to telecommute is consistent with numbers reported elsewhere (Dingel and Neiman, 2020; Parker et al., 2020).

The fraction of workers who *choose* not to telecommute dropped to near zero during the pandemic, and is expected to stabilize at about half of its pre-pandemic level. Of those employed, more than half reported telecommuting sometimes or every day during the pandemic, consistent with others' findings (Barrero et al., 2021; Brynjolfsson et al., 2020). This is more than twice the fraction of the workforce that telecommuted pre-COVID, and most of these workers plan to continue telecommuting at least some of the time post-pandemic. Our estimate that around 35% of workers were fully remote during the pandemic is also consistent with other estimates from May 2020 (Barrero et al., 2021; Bick et al., 2021).

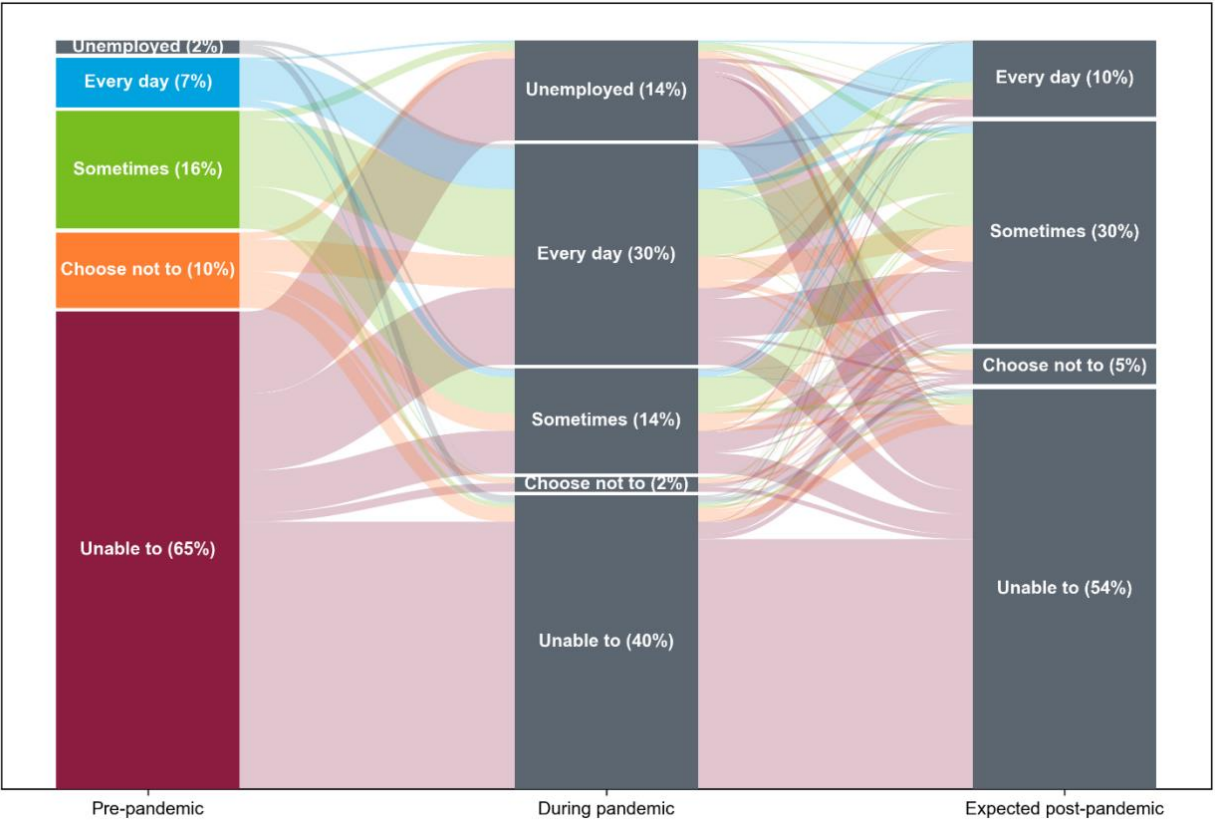

**Figure 1: Sankey Plot of Weighted Telecommuting Outcomes in each Period, Wave 1, COVID Future Panel Survey**

Looking at the flows between periods is informative as well. Those who became unemployed during the pandemic were dominated by workers who were unable to telecommute, and most of them expect to return to jobs where they will be unable to telecommute. Of the workers who expect to telecommute sometimes post-COVID, about half of them did not telecommute pre-COVID. Of the workers who expect to telecommute every day, about one quarter of them did not telecommute pre-COVID. Overall, comparing their pre- and post-pandemic telecommuting patterns in Wave 1, we find that more than 20% of workers expect to telecommute more, 3% expect to telecommute less, and the remaining 75% expect no change. By Wave 3, this comparison becomes more than 30% of workers expecting to telecommute more, and 66% expecting no change.

**Stickiness of the pandemic telecommuting experience**

The COVID pandemic gave a lot of workers their first major experience with telecommuting, and there are good reasons to believe these experiences could be "sticky." Namely, telecommuting became both more effective and normalized, workers might enjoy the flexibility of remote work, and employers might prefer remote employees to reduce overhead costs. As long as productivity does not drop, retaining the pandemic-era expansion of remote work could be a win-win corporate policy.

Another informative way to look at the COVID Future Panel Survey data, then, is to divide workers into categories based on their telecommuting experience pre-COVID and during the pandemic, and see what these groups with different experiences expect for their futures – especially focusing on the new telecommuters. **Table 1** does this, dividing workers into three main experience categories, and subdividing the New Telecommuter category into two subcategories.

The first thing to note is the high percentage of workers in every experience category who expect to telecommute at least sometimes post-COVID if they are given the option (see the Telecommute if Able row). This corresponds to the small fraction of workers who choose not to telecommute from **Figure 1**, and clearly indicates that telecommuting is popular among workers.

**Table 1: Weighted Post-COVID Telecommuting Expectations by Telecommuting Experience Categories, Wave 1**

| | Telecommuted Pre-COVID | New Telecommuter | | Never Telecommuted |
| | | Chose Not to Telecommute Pre-COVID | Unable to Telecommute Pre-COVID | |
| **Wave 1 Post-COVID Expectation** | | | | |
| Unable to Telecommute | 6% | 13% | 36% | 86% |
| Choose Not to Telecommute | 2% | 17% | 4% | 5% |
| Telecommute Sometimes | 59% | 65% | 49% | 8% |
| Telecommute Every Day | 32% | 5% | 11% | 2% |
| Telecommute if Able | 98% | 80% | 93% | 67% |
| Weighted % of all Workers | 23% | 7% | 17% | 53% |
| N (unweighted) | 1,337 | 387 | 948 | 2,224 |

Of the new pandemic telecommuters, about 70% were unable to work from home before the pandemic, but the remaining 30% had previously chosen not to do so. As is clear from **Table 1**, it is these new telecommuters who expect the largest shift in their post-COVID telecommuting frequencies, relative to their pre-COVID baseline. Despite the fact that none of these workers telecommuted before the pandemic, more than 60% of them expect to telecommute sometimes or every day after the pandemic.

Focusing on the new telecommuters who gained telecommuting flexibility during the pandemic *and* expect to have the option to continue the practice post-COVID, we find that telecommuting is a highly sticky choice; nearly *all* of them (93%) expect to telecommute at least sometimes, and more than 15% expect to telecommute every day. Among those pandemic telecommuters who chose not to telecommute pre-COVID and expect to have the option to choose again post-COVID, 80% expect to switch to telecommuting at least sometimes.

Data from Wave 2 and Wave 3 of the COVID Future Panel Survey produce the same basic patterns; virtually all of the telecommuting expectations are stable or moving toward *more* telecommuting. Notably, in Wave 3, fewer respondents report that they are unable to telecommute or might choose not to do so, and more respondents report that they expect to telecommute sometimes or every day. See Appendix **Table A1** and **Table A2** for details.

**Who are the new pandemic telecommuters?**

We now know that most of the new pandemic telecommuters plan to continue telecommuting if they have the opportunity. This finding makes it important to know who these new telecommuters are, and whether they differ in key ways from pre-pandemic telecommuters.

**Figure 2** illustrates the sociodemographic composition of workers in the three telecommuting experience categories: Telecommuted Pre-COVID, New Telecommuter, and Never Telecommuted. Across multiple dimensions, new telecommuters are remarkably similar to those who telecommuted before the pandemic and quite different from those who never telecommuted. Compared to workers who never telecommuted, they have high educational attainment and income, are more likely to be non-Hispanic white and less likely to be Hispanic, and are more likely to be in the 35-49 age group. In particular, education is often considered a pathway to achieving career benefits such as greater pay and higher job security (Neumark, 2000; Turner et al., 2007), but fringe benefits such as teleworking also appear to be important opportunities primarily available to highly educated workers. Although some literature from before COVID-19 indicates that gender, race, and age are important determinants of telecommuting (Georgiana, 2016; Walls et al., 2007), **Figure 2** demonstrates that these demographics display relatively weak relationships with telecommuting experience in the COVID Future data.

Figure 3 illustrates the composition of workers in the three telecommuting experience categories along other dimensions relating to work. Notably, the distribution of job types among new telecommuters is much more similar to that among pre-COVID telecommuters than that among those who never telecommuted. The category of frontline workers (healthcare, emergency response, transportation, maintenance, grocery, and hospitality) is much larger in the "Never Telecommuted" category. This category contains a high percentage of jobs which cannot be done from home. On the other hand, workers in the education sector experienced a mass transition to online schooling during the pandemic, with this industry being overrepresented among new telecommuters. Professional workers, many of whom have jobs that could be done remotely, enjoyed high rates of telecommuting before the pandemic. These workers also experienced substantial growth in telecommuting during the pandemic.

In addition, **Figure 3** illustrates that new telecommuters were less likely to report stable productivity in their jobs, and slightly more likely to report increased productivity. This productivity difference disappears by Wave 3 of the COVID Future Panel Survey, however. Although these are self-reported survey responses, this finding is promising for a future of increased remote work.

# Educational Attainment

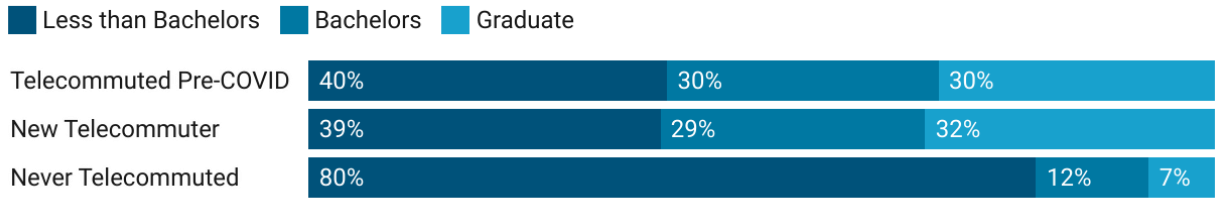

| | Less than Bachelors | Bachelors | Graduate |
|---|---|---|---|
| Telecommuted Pre-COVID | 40% | 30% | 30% |
| New Telecommuter | 39% | 29% | 32% |
| Never Telecommuted | 80% | 12% | 7% |

# Income

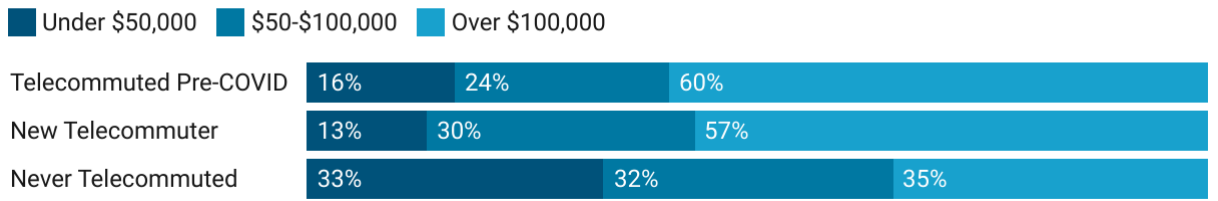

| | Under $50,000 | $50-$100,000 | Over $100,000 |
|---|---|---|---|
| Telecommuted Pre-COVID | 16% | 24% | 60% |
| New Telecommuter | 13% | 30% | 57% |
| Never Telecommuted | 33% | 32% | 35% |

# Race/Ethnicity

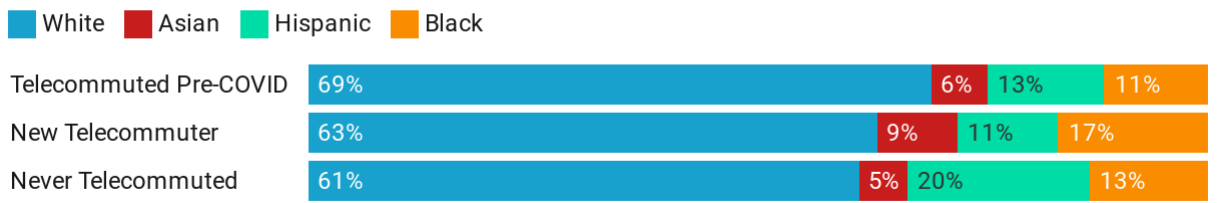

| | White | Asian | Hispanic | Black |
|---|---|---|---|---|
| Telecommuted Pre-COVID | 69% | 6% | 13% | 11% |
| New Telecommuter | 63% | 9% | 11% | 17% |
| Never Telecommuted | 61% | 5% | 20% | 13% |

# Gender + Children

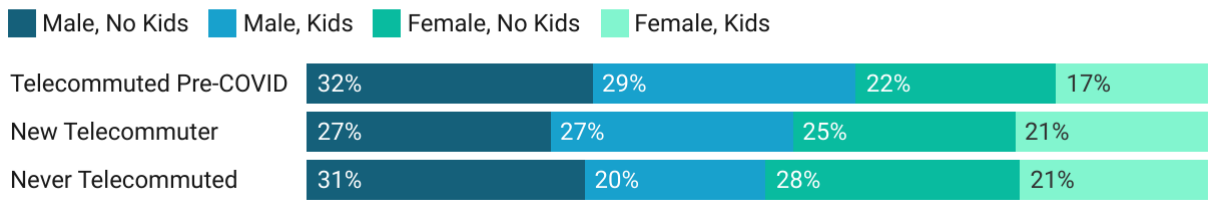

| | Male, No Kids | Male, Kids | Female, No Kids | Female, Kids |
|---|---|---|---|---|
| Telecommuted Pre-COVID | 32% | 29% | 22% | 17% |
| New Telecommuter | 27% | 27% | 25% | 21% |
| Never Telecommuted | 31% | 20% | 28% | 21% |

# Age

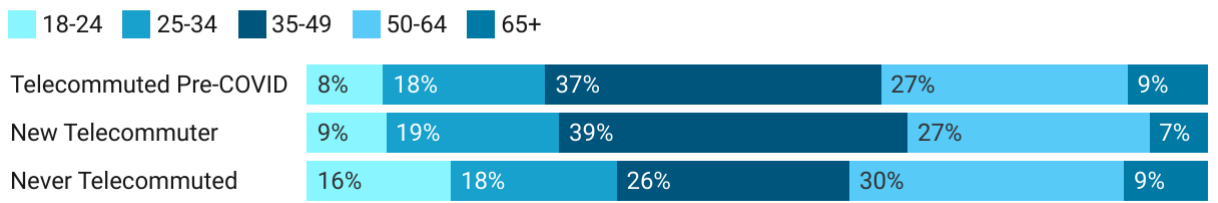

| | 18-24 | 25-34 | 35-49 | 50-64 | 65+ |
|---|---|---|---|---|---|
| Telecommuted Pre-COVID | 8% | 18% | 37% | 27% | 9% |
| New Telecommuter | 9% | 19% | 39% | 27% | 7% |
| Never Telecommuted | 16% | 18% | 26% | 30% | 9% |

**Figure 2: Weighted sociodemographic distributions by telecommuter experience category, Wave 1**

# Job Category

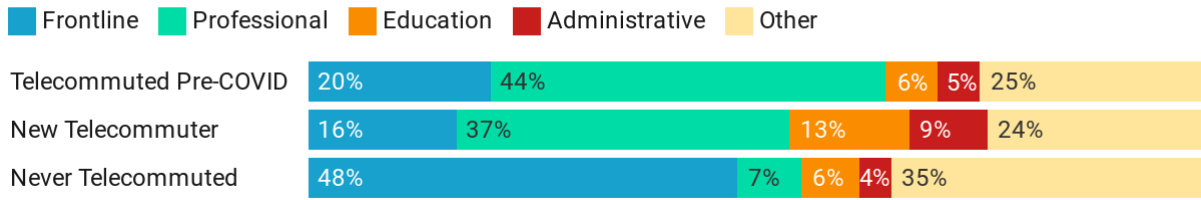

# Pre-COVID Commute Mode

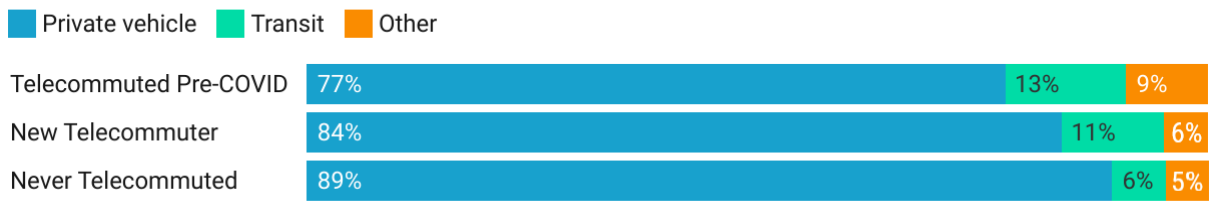

# COVID Productivity Change

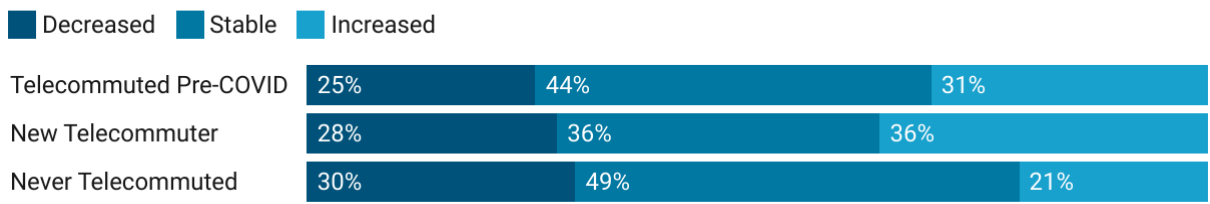

# Like Working from Home

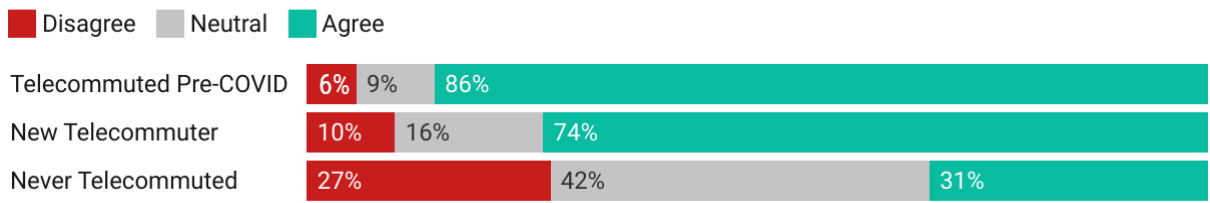

# Hard to Motivate Away from the Office

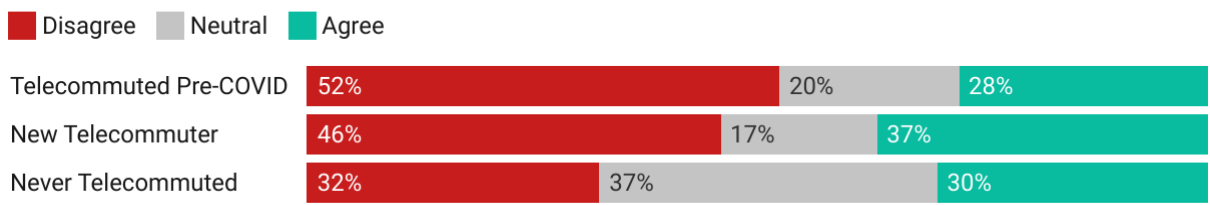

**Figure 3: Weighted distributions of work-related characteristics by telecommuter experience category, Wave 1**

flexibility without largescale productivity loss. The two attitudinal variables included in **Figure 3** also help to clarify opportunities and challenges in a future of expanded remote work. Those with experience telecommuting overwhelmingly agree with the statement, "I like working from home," suggesting that quality of life for workers may be higher. Especially those new to telecommuting also often agree with the statement, "It is hard to get motivated to work away from the main office," suggesting that for many workers, full-time remote work may not be ideal. For both of these variables, those who never telecommuted have answers that are evenly distributed around neutral. This is likely because without telecommuting experience, it is difficult to reliably report attitudes about telecommuting. Except where noted, findings reported in **Figure 2** and **Figure 3** are unchanged in Waves 2 and 3 of the COVID Future Panel Survey.

**Determinants of telecommuting in three periods**

So far, we have illustrated two main findings. First, a lot of workers experienced telecommuting during the pandemic and plan to continue telecommuting at least part time going forward. Second, new pandemic telecommuters are similar in many ways to those who telecommuted pre-COVID.

Here, we employ multivariate statistical analysis to identify the contributions of a variety of factors to telecommuting outcomes in three periods: pre-COVID, the pandemic period, and expectations for the post-COVID future. The results tell us which factors influence telecommuting in each period, the estimated magnitudes of these relationships, and whether and how much this influence has changed across these three time periods.

*A joint model of telecommuting ability and frequency*

Telecommuting is a combination of two related but distinct outcomes—having a job that allows it and choosing to take advantage of that ability either part-time or full-time (Singh et al., 2013). Since the factors determining the ability to telecommute are likely somewhat different from those determining telecommuting frequency, we model the option to telecommute separately from the choice of telecommuting frequency for workers who have the option.

Estimating separate models presents a sample selection problem in the telecommuting frequency model, however. We are interested in modeling the choice process for how workers decide whether and how often they telecommute across the entire worker population, but we only observe telecommuting frequency for workers who have the option to telecommute. The portion of the working population that has the option to telecommute is likely different from the overall worker population in ways that affect the choice of telecommuting frequency, but that we do not observe and therefore cannot control for in our model. If this is true and left uncorrected, results from the telecommuting frequency model will be biased. We correct this bias by jointly estimating a multinomial probit model of telecommuting frequency and a binary probit model of the ability to telecommute, using the -cmp- package in Stata (Roodman, 2011). For additional details about our estimation strategy, please see the Supplemental Materials linked to this article.

In order to select predictor variables to include in our joint model of telecommuting ability and frequency, we developed a conceptual model of the determinants of telecommuting (**Figure 4**). The two dependent variables are in boldface in the central two boxes, predictor variables are in boxes surrounding them, arrows indicate connections between variables and directions of influence, and box shading indicates variables that the COVID Future survey does not measure. As is evident from **Figure 4**, most predictor variables affect either the ability to telecommute or telecommuting frequency, but not both.

We posit that whether a worker has the ability to telecommute is the result of a "negotiation" between the worker and their direct supervisor. Often, an actual negotiation never occurs. This could be because the worker's job duties prohibit or force telecommuting, the employer's telecommuting policy dictates the outcome, or because neither the worker nor their supervisor is interested in telecommuting. Where a negotiation does take place, the outcome is influenced by who the worker is (higher socioeconomic status workers likely have more bargaining power), their experience on the job (more experience leads to greater worker-supervisor trust), actual job duties (some jobs require at least part-time physical presence at the workplace, while others do not), and the worker's productivity while telecommuting.

Telecommuting frequency is the choice by the worker of how often to telecommute if they are able. While this choice is influenced somewhat by employer policy, we conceptualize the choice of telecommuting frequency as being made chiefly by the worker. As such, most of the predictors of telecommuting frequency relate to the worker's experience of telecommuting and the worker's experience of commuting to and working at the workplace. The overarching hypothesis that drives the relationships depicted in **Figure 4** is that workers who have unpleasant and/or long commutes or find the workplace to be unsafe or unconducive to productive work will telecommute more, as will workers who enjoy telecommuting and have a good home setup for work.

Since the onset of the COVID-19 pandemic, workers in the U.S. have actually gained power relative to their employers. After an initial dip, there was a strong labor market in which many workers switched jobs to gain higher pay and/or a better working situation for themselves – including more flexible working hours and location. For this reason, we expect that even if employers would prefer to have employees in the office full time, they may compromise to allow at least some telecommuting in order to retain their workforce.

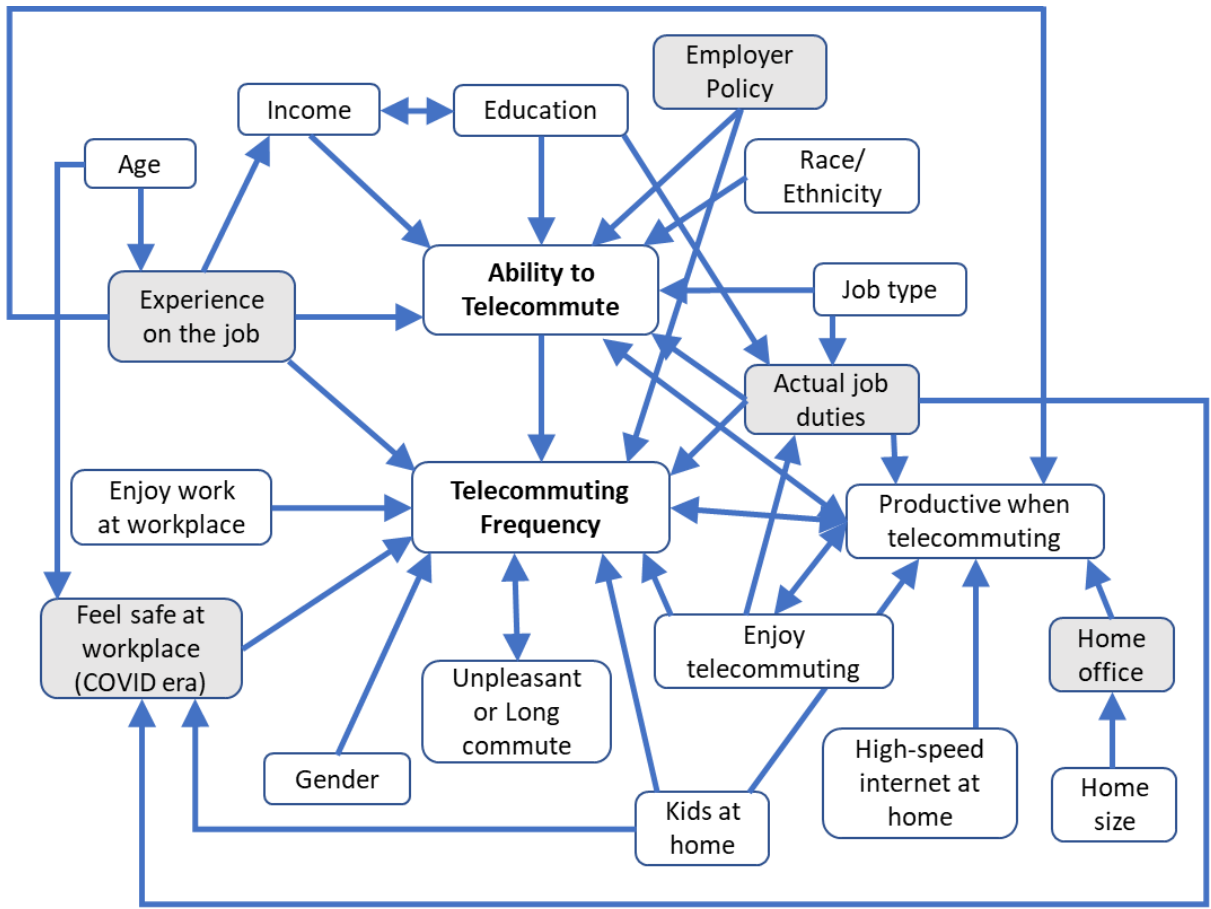

**Figure 4: Conceptual model of the determinants of telecommuting**

The estimated models include all but one of the variables in **Figure 4** that are in unshaded boxes. The one exception is productivity while telecommuting. We do not include this variable because we think that it is not only a predictor of telecommuting, but that it may also be predicted by telecommuting (i.e. the more a worker telecommutes, the more productive they are while telecommuting). Instead, the model includes a number of variables that we expect relate to productivity while telecommuting (i.e. kids at home, age, home size, job type, internet at home, education level, and enjoyment of telecommuting).

In addition to the variables depicted in **Figure 4**, we also tested the direct effect of the pandemic on telecommuting by including variables representing county-level COVID restriction policies that were in effect (stay-at-home orders, mask mandates, and bar and restaurant closures and restrictions), as well as county-level COVID case rates at the time that respondents took the survey. The restrictions were not statistically significant in any period, and the COVID case rates had an extremely small positive effect on telecommuting ability in the "during COVID" period using the Wave 1 dataset. Because our main focus is the effect of the pandemic experience on the future of telecommuting, however, our final models do not include this variable.

It is interesting and somewhat surprising that COVID-related government restrictions, which did vary widely across our sample in both space and time, were unrelated to telecommuting outcomes even during the pandemic. This may be because most people who could easily telecommute were doing so during much of this period, and local government policies did not affect their choices.

We report our results as marginal effects in **Table 2** and **Table 3**. Although the models were estimated without weights, these marginal effects are calculated with weights to represent the average effect across the population. These marginal effects represent the predicted change in the probability of each outcome resulting from a one-unit increase in the variable of interest. For instance, controlling for all other variables in the model, compared with a worker who does not hold a bachelor's degree, a graduate degree holder has a predicted probability of having the option to telecommute that is 0.18 (18 percentage points) higher in the pre-pandemic period, 0.26 (26 percentage points) higher during the pandemic, and 0.20 (20 percentage points) higher in the post-pandemic period.

Education level and job category are strongly associated with the ability to telecommute in all periods. Workers with bachelor's and especially graduate degrees are much more likely have the option to telecommute than those without. Professional workers are more likely to have the option to telecommute than the general category of "Other" workers, while frontline workers are less likely to have the option. Those who work in the education sector are less likely than "Other" workers to have the option to telecommute in the pre- and post-pandemic periods, but during the pandemic are more likely to—reflecting the fact that much of the education sector moved online during the pandemic. Data from Wave 2 (collected mainly between November 2020 and March 2021) and Wave 3 (collected mainly in November 2021), suggest that during those periods, education workers were equally as likely as "Other" workers to have the option to telecommute.

Workers living in dense urban centers were estimated to be more likely to have the option to telecommute in the Wave 1 data in all periods, but the Wave 2 and Wave 3 analysis suggests that this urban effect might disappear in the post-COVID period. Workers in households with incomes over $100,000 are more likely to have the option as well. Workers who changed jobs since the start of the pandemic were more likely to expect to be able to telecommute post-pandemic, suggesting workers may be valuing this option in their job search. This relationship became statistically insignificant in the Wave 2 and Wave 3 analyses. In all periods, employed students were more likely to have the option to telecommute for their jobs. This student effect persisted in the Wave 2 data, but disappeared in Wave 3.

Frequency of telecommuting is modeled as the choice between never or rarely telecommuting (i.e. a few times per year), telecommuting sometimes (i.e. between a few times per month and a few times per week), and telecommuting every day. For each predictor variable included in the model, we report the weighted marginal effect of a one unit increase on the probability of a worker choosing each of these telecommuting frequency options. The three marginal effects sum to 0 (though this is not always obvious in **Table 3** due to rounding).

Workers who like working from home, unsurprisingly, choose to do it more. Workers who enjoy workplace interaction are slightly less likely to choose to telecommute every day, as are workers who find it hard to motivate away from the office. Note that these attitudinal variables are measured on a 5-point Likert scale, meaning that the reported marginal effects should be interpreted as the effect associated with a one-point change in that scale. For instance, the effect of changing from an attitude of "Somewhat Agree" to "Strongly Agree" with the statement "I like working from home" is associated with a 5 percentage point increase in the likelihood of telecommuting every day in the post-COVID period, and a 5 percentage point decrease in the likelihood of never or rarely telecommuting.

Turning to other predictors, those who commuted by transit pre-pandemic are likely to telecommute sometimes after the pandemic, but not every day. With age, workers are increasingly likely to telecommute every day, and less likely to choose to never or rarely telecommute. Somewhat surprisingly, home factors that we thought would affect telecommuting (high speed internet, extra bedroom, children, and urban resident) were not important predictors of telecommuting in any period. In addition, controls for region of the U.S. (estimates not shown in **Table 2** and **Table 3**) indicate again that telecommuting outcomes do not exhibit strong spatial patterns.

Comparing Wave 2 and 3 marginal effect estimates for telecommuting frequency with those in Table 3, we find almost complete agreement. The two differences are both in the post-COVID expectations from the Wave 3 model; the effects of enjoying workplace interaction become larger, and the effect of enjoying telecommuting on telecommuting every day also becomes larger.

**Table 2:** Weighted Marginal Effects on the Option to Telecommute in Three Periods, Wave 1

| | Pre-pandemic | S.E. | During pandemic | S.E. | Expected Post-pandemic | S.E. |
|---|---|---|---|---|---|---|
| *Educational attainment* | | | | | | |
| Less than bachelors | base | | base | | base | |
| Bachelors | 0.10*** | 0.02 | 0.15*** | 0.02 | 0.13*** | 0.02 |
| Graduate | 0.18*** | 0.02 | 0.26*** | 0.02 | 0.20*** | 0.02 |
| *Job category* | | | | | | |
| Other | base | | base | | base | |
| Frontline | -0.17*** | 0.02 | -0.23*** | 0.02 | -0.17*** | 0.02 |
| Professional | 0.13*** | 0.02 | 0.24*** | 0.02 | 0.24*** | 0.02 |
| Education | -0.14*** | 0.02 | 0.09*** | 0.03 | -0.13*** | 0.02 |
| Administrative | NS | | 0.19*** | 0.03 | 0.10*** | 0.03 |
| *Household income* | | | | | | |
| Under $50,000 | base | | base | | base | |
| $50-$100,000 | NS | | 0.03* | 0.02 | 0.01 | 0.02 |
| Over $100,000 | 0.10*** | 0.02 | 0.13*** | 0.02 | 0.12*** | 0.02 |
| *Race/Ethnicity* | | | | | | |
| White/Other | base | | base | | base | |
| Hispanic | NS | | NS | | NS | |
| Black | 0.04* | 0.02 | 0.05** | 0.02 | 0.04** | 0.02 |
| Asian | NS | | NS | | NS | |
| *Age category* | | | | | | |
| 18-24 | base | | base | | base | |
| 25-34 | 0.07*** | 0.03 | NS | | NS | |
| 35-49 | 0.05* | 0.03 | NS | | NS | |
| 50-64 | 0.06** | 0.03 | NS | | NS | |
| 65+ | 0.08** | 0.03 | NS | | NS | |
| Student | 0.12*** | 0.02 | 0.12*** | 0.02 | 0.11*** | 0.02 |
| Dense urban resident | 0.06*** | 0.02 | 0.10*** | 0.02 | 0.06*** | 0.02 |
| Full time worker | NS | | NS | | -0.03* | 0.02 |
| Job change | NA | | NS | | 0.05** | 0.02 |
| | | | | | | |
| Observations | 4,766 | | 4,163 | | 4,854 | |

NS = "Not Significant," NA = "Not Applicable," Region of US controlled in model
*** p<0.01, ** p<0.05, * p<0.1

**Table 3:** Weighted Marginal Effects on Telecommuting Frequency in Three Periods, Wave 1

| | Pre-pandemic | S.E. | During pandemic | S.E. | Expected Post-pandemic | S.E. |
|---|---|---|---|---|---|---|
| **Student** | | | | | | |
| Never/Rarely Telecommute | -0.13*** | 0.04 | -0.01 | 0.01 | NS | |
| Sometimes Telecommute | 0.10** | 0.04 | 0.11*** | 0.03 | NS | |
| Telecommute Every Day | 0.03** | 0.01 | -0.10*** | 0.03 | NS | |
| *Job Category* | | | | | | |
| Other | base | | base | | base | |
| Frontline | | | | | | |
| Never/Rarely Telecommute | 0.01 | 0.04 | -0.01 | 0.01 | -0.01 | 0.02 |
| Sometimes Telecommute | 0.04 | 0.04 | 0.11*** | 0.04 | 0.04** | 0.02 |
| Telecommute Every Day | -0.05*** | 0.01 | -0.09*** | 0.04 | -0.03*** | 0.01 |
| Professional | | | | | | |
| Never/Rarely Telecommute | NS | | 0.02 | 0.01 | 0.00 | 0.02 |
| Sometimes Telecommute | NS | | -0.08** | 0.03 | -0.03 | 0.02 |
| Telecommute Every Day | NS | | 0.06** | 0.03 | 0.04*** | 0.01 |
| Education | | | | | | |
| Never/Rarely Telecommute | 0.05 | 0.04 | 0.02* | 0.01 | 0.02 | 0.02 |
| Sometimes Telecommute | -0.00 | 0.04 | -0.04 | 0.03 | 0.04 | 0.03 |
| Telecommute Every Day | -0.05*** | 0.02 | 0.02 | 0.03 | -0.05*** | 0.01 |
| Administrative | | | | | | |
| Never/Rarely Telecommute | NS | | NS | | NS | |
| Sometimes Telecommute | NS | | NS | | NS | |
| Telecommute Every Day | NS | | NS | | NS | |
| *Attitudes* | | | | | | |
| Enjoy Workplace Interaction | | | | | | |
| Never/Rarely Telecommute | 0.02 | 0.01 | 0.00 | 0.00 | 0.00 | 0.01 |
| Sometimes Telecommute | 0.01 | 0.01 | 0.03*** | 0.01 | 0.03*** | 0.01 |
| Telecommute Every Day | -0.03*** | 0.01 | -0.03*** | 0.01 | -0.02*** | 0.00 |
| Hard to Motivate | | | | | | |
| Never/Rarely Telecommute | 0.02 | 0.01 | 0.00 | 0.00 | 0.02** | 0.01 |
| Sometimes Telecommute | 0.00 | 0.01 | 0.02*** | 0.01 | 0.00 | 0.01 |
| Telecommute Every Day | -0.01** | 0.00 | -0.02*** | 0.01 | -0.02*** | 0.00 |
| Enjoy Telecommuting | | | | | | |
| Never/Rarely Telecommute | -0.10*** | 0.01 | -0.03*** | 0.01 | -0.05*** | 0.01 |
| Sometimes Telecommute | 0.06*** | 0.02 | -0.04*** | 0.01 | 0.01 | 0.01 |
| Telecommute Every Day | 0.04*** | 0.01 | 0.07*** | 0.01 | 0.05*** | 0.01 |
| Dense Urban Resident | | | | | | |
| Never/Rarely Telecommute | NS | | NS | | NS | |
| Sometimes Telecommute | NS | | NS | | NS | |
| Telecommute Every Day | NS | | NS | | NS | |

| | Pre-pandemic | S.E. | During pandemic | S.E. | Expected Post-pandemic | S.E. |
|---|---|---|---|---|---|---|
| High Speed Internet at Home | | | | | | |
|   Never/Rarely Telecommute | NS | | NS | | NS | |
|   Sometimes Telecommute | NS | | NS | | NS | |
|   Telecommute Every Day | NS | | NS | | NS | |
| Extra Bedroom at Home | | | | | | |
|   Never/Rarely Telecommute | NS | | NS | | NS | |
|   Sometimes Telecommute | NS | | NS | | NS | |
|   Telecommute Every Day | NS | | NS | | NS | |
| Female | | | | | | |
|   Never/Rarely Telecommute | NS | | 0.01 | 0.01 | NS | |
|   Sometimes Telecommute | NS | | -0.07*** | 0.02 | NS | |
|   Telecommute Every Day | NS | | 0.07*** | 0.02 | NS | |
| Children | | | | | | |
|   Never/Rarely Telecommute | NS | | NS | | 0.00 | 0.02 |
|   Sometimes Telecommute | NS | | NS | | 0.02 | 0.02 |
|   Telecommute Every Day | NS | | NS | | -0.02** | 0.01 |
| *Race/Ethnicity* | | | | | | |
| White/Other | base | | base | | base | |
| Hispanic | | | | | | |
|   Never/Rarely Telecommute | NS | | NS | | NS | |
|   Sometimes Telecommute | NS | | NS | | NS | |
|   Telecommute Every Day | NS | | NS | | NS | |
| Black | | | | | | |
|   Never/Rarely Telecommute | 0.06 | 0.04 | 0.01 | 0.01 | NS | |
|   Sometimes Telecommute | -0.07* | 0.04 | -0.08** | 0.03 | NS | |
|   Telecommute Every Day | 0.01 | 0.02 | 0.07** | 0.03 | NS | |
| Asian | | | | | | |
|   Never/Rarely Telecommute | 0.03 | 0.04 | NS | | NS | |
|   Sometimes Telecommute | 0.02 | 0.04 | NS | | NS | |
|   Telecommute Every Day | -0.05** | 0.02 | NS | | NS | |
| *Age category* | | | | | | |
| 18-24 | base | | base | | base | |
| 25-34 | | | | | | |
|   Never/Rarely Telecommute | NS | | NS | | -0.01 | 0.03 |
|   Sometimes Telecommute | NS | | NS | | -0.02 | 0.03 |
|   Telecommute Every Day | NS | | NS | | 0.04*** | 0.01 |
| 35-49 | | | | | | |
|   Never/Rarely Telecommute | -0.08 | 0.06 | NS | | -0.04 | 0.03 |
|   Sometimes Telecommute | 0.03 | 0.06 | NS | | -0.01 | 0.03 |
|   Telecommute Every Day | 0.05*** | 0.02 | NS | | 0.06*** | 0.01 |
| 50-64 | | | | | | |
|   Never/Rarely Telecommute | -0.08 | 0.06 | NS | | -0.06* | 0.03 |
|   Sometimes Telecommute | 0.01 | 0.06 | NS | | 0.00 | 0.03 |

| | Pre-pandemic | S.E. | During pandemic | S.E. | Expected Post-pandemic | S.E. |
|---|---|---|---|---|---|---|
| Telecommute Every Day | 0.07*** | 0.02 | NS | | 0.06*** | 0.02 |
| 65+ | | | | | | |
|   Never/Rarely Telecommute | -0.14** | 0.07 | NS | | -0.10*** | 0.03 |
|   Sometimes Telecommute | 0.03 | 0.07 | NS | | 0.00 | 0.04 |
|   Telecommute Every Day | 0.11*** | 0.03 | NS | | 0.10*** | 0.02 |
| *Job/Commute Characteristics* | | | | | | |
| Nonmotorized commuter pre-pandemic | | | | | | |
|   Never/Rarely Telecommute | NS | | 0.01 | 0.02 | NS | |
|   Sometimes Telecommute | NS | | 0.11*** | 0.04 | NS | |
|   Telecommute Every Day | NS | | -0.11*** | 0.05 | NS | |
| Transit commuter pre-pandemic | | | | | | |
|   Never/Rarely Telecommute | -0.02 | 0.04 | NS | | -0.06** | 0.03 |
|   Sometimes Telecommute | 0.12*** | 0.04 | NS | | 0.10*** | 0.03 |
|   Telecommute Every Day | -0.13*** | 0.04 | NS | | -0.04*** | 0.02 |
| Full time | | | | | | |
|   Never/Rarely Telecommute | 0.13*** | 0.03 | 0.00 | 0.01 | NS | |
|   Sometimes Telecommute | -0.08** | 0.03 | -0.15*** | 0.03 | NS | |
|   Telecommute Every Day | -0.05*** | 0.02 | 0.15*** | 0.03 | NS | |
| Job Change | | | | | | |
|   Never/Rarely Telecommute | NA | | -0.02 | 0.01 | NS | |
|   Sometimes Telecommute | NA | | 0.07** | 0.03 | NS | |
|   Telecommute Every Day | NA | | -0.05* | 0.03 | NS | |
| | | | | | | |
| Observations | 4,766 | | 4,163 | | 4, 854 | |

NS = "Not Significant," NA = "Not Applicable," Region of US controlled in model
*** p<0.01, ** p<0.05, * p<0.1

The results of our analysis largely concur with the literature on this topic. Although we found full-time workers were more likely to be completely remote during the pandemic, these respondents were less likely to be fully remote pre-pandemic, which others have found (Sener and Bhat, 2011; Tang et al., 2011; Walls et al., 2007). Our finding that urbanites were more likely to have the option to telecommute is also consistent with other research (Singh et al., 2013). Furthermore, the positive relationship between age and telecommuting found in our data is well-established by others (Asgari et al., 2014; Sener and Bhat, 2011; Walls et al., 2007).

Although the positive relationship between income and ability to telecommute is consistently reported by researchers (Sener and Bhat, 2011; Tang et al., 2011), the impact of educational attainment on the ability to telecommute is confirmed by some (Sener and Bhat, 2011; Tang et al., 2011; Walls et al., 2007) but not all (Singh et al., 2013). Similarly, some research has found industry to be unrelated to the ability to telecommute (Singh et al., 2013), although most find a relationship (Asgari et al., 2014; Sener and Bhat, 2011; Walls et al., 2007).

### *Stability across periods in the determinants of telecommuting*

There are almost no statistically significant differences in the predictors of telecommuting before and after the pandemic; the confidence intervals of almost all marginal effects overlap. In the Wave 1 analysis, there were three relationships that changed. First, professional, managerial, technical, and administrative workers (labeled "Professional" and "Administrative" in **Table 2**) are significantly more likely to have the option to telecommute post-pandemic than they did pre-pandemic. Second, the level of agreement with the statement "I like working from home" is less negatively associated with never or rarely telecommuting in the post-COVID period than in the pre-COVID period. Third, full time worker status is no longer positively associated with never or rarely telecommuting in the post-COVID period.

In the Wave 2 and 3 analyses, the confidence intervals for all marginal effect estimates are much larger because the sample sizes were much smaller. Therefore, even where point estimates diverge substantially between periods, their confidence intervals usually overlap. The only places where we find a statistically significant difference between the pre- and post-COVID models are related to attitudinal statements in the Wave 3 analysis. Specifically, the positive association between level of agreement with the statement "I like working from home" and telecommuting every day became stronger in the post-COVID period, and those who enjoy social interaction at their workplace become more likely to telecommute sometimes.

There are larger differences during the pandemic. In particular, in Wave 1, education workers are much more likely to have the option to telecommute during the pandemic than in any other period, reflecting the mass movement of schooling online. During the period that we fielded Waves 2 and 3, many schools had returned to in-person instruction, so this result disappears from our models as well. We still find in Wave 2, however, that if offered the option at that time, education workers were still much more likely to choose to telecommute every day than they were pre-COVID. Full-time workers were much more likely to choose to telecommute every day during the pandemic than before or after in all waves of our survey. In Waves 2 and 3, we also find that workers who were transit commuters pre-COVID were likely to choose every day telecommuting in the survey period, but to expect to telecommute only sometimes in the post-COVID future. In Wave 3, we also find that those who enjoy interaction at their workplace are much more likely to choose to telecommute sometimes, compared to their choices when they took the survey.

It is striking that age does not appear to influence telecommuting during the pandemic, especially because age is so closely related to severe disease from COVID-19. There are two main explanations for this counterintuitive finding. First, the vast majority of workers who were able to telecommute during the pandemic did so, in all age categories. Second, a large fraction of workers in the oldest age category (65+) stopped working during the pandemic and are therefore not included in the during pandemic analysis.

**DISCUSSION**

Perhaps the single most important conclusion from the COVID Future Panel Survey is that we should anticipate a long-term increase in telecommuting among US workers. Here, we put together key summary statistics from the survey to fully explain this conclusion. For all percentages in these paragraphs, we provide the Wave 1 result, immediately followed by the Wave 2 and Wave 3 results in parentheses. In all cases, Wave 2 and 3 data provide even stronger evidence for the projected long-term increase in telecommuting; people's expectations for the future are not reverting to what was the pre-COVID norm.

The fraction of workers who telecommute at least a few times per month is projected to increase from 23% to 40% (W2:41%, W3:49%). The projected increase in those who telecommute at least a few times per week is even steeper, from 13% to 26% (W2:28%, W3:36%) of workers.

Part of this projected increase stems from expectations from more workers that they will have the option to work remotely. The fraction of US workers who expect to have the option to telecommute is 11 (W2:9, W3:19) percentage points higher than it was in the pre-COVID era.

There is also a clear increase in worker appetite for telecommuting, however. Among those workers who had the option to telecommute pre-COVID and expect to have the option in the future (nearly 30% (W2:26%, W3:33%) of workers), almost 25% (W2:39%, W3:43%) expect to increase their telecommuting frequency long-term. Among workers who had their first experience with regular telecommuting during the pandemic (more than 20% (W2:27%, W3:29%) of workers), 61% (W2:65%, W3:78%) of them expect to continue to telecommute. Among the 10% (W2:11%, W3:16%) of workers who gained both the ability and the experience of telecommuting during the pandemic and expect to have the option to continue, 93% (W2:94%, W3:97%) expect to do so. Nearly 20% (W2:19%, W3:30%) of these workers expect to telecommute *every day*.

This increase in telecommuting seems to be a welcome new reality for most. New telecommuters especially appreciate the quality of life benefits of telecommuting, with two-thirds of them in all waves reporting that telecommuting some of the time and/or commuting less are among the top three aspects of pandemic life that they would like to extend post-pandemic. Three-quarters (W2:79%, W3:75%) of new and more than 85% (W2:89%, W3:86%) of experienced telecommuters agreed with the attitudinal statement "I like working from home," while those without telecommuting experience responded neutrally to this statement. Taken together, our data strongly suggest that for those who have experienced telecommuting, at least part-time remote work is something that most would like to continue post-pandemic. This finding is consistent with reports from other pandemic-era surveys. Pew Research Center's October 2020 survey found that among those who can do their jobs remotely, nearly 90% would like to telecommute at least some of the time post-pandemic (Parker et al., 2020).

A question that often arises is whether telecommuting affects worker productivity (Allen et al., 2015). A substantial body of literature has been devoted to identifying factors which tend to boost the productivity of remote workers. Thorough communication, support with technology, assignment of skilled and creative tasks, and freedom to live beyond commuting distance from the office all have been found to increase productivity gains among remote workers (Choudhury et al., 2021; Dutcher, 2012; Kemerling, 2002; Monteiro et al., 2019).

The COVID Future survey suggests three encouraging findings regarding the productivity outcomes of telecommuters. First, nearly three-quarters (W2:78%, W3:89%) of workers reported stable or increased productivity since the start of the pandemic. Second, the fraction of workers in Wave 1 reporting an increase in their productivity was highest for workers who were new to telecommuting. This difference disappeared in Waves 2 and 3, though telecommuters in all waves were more likely to report an increase in productivity than non-telecommuters. Third, productivity was especially correlated with post-pandemic telecommuting expected frequency for those workers who were new telecommuters during the pandemic. More than 80% (W2:80%, W3:90%) of new telecommuters who expect to continue telecommuting frequently reported stable or increased productivity, and more than 80% (W2:80%, W3:70%) of new telecommuters who expect not to continue telecommuting reported stable or decreased productivity. If remote work does improve overall worker output, especially once pandemic-related stresses and challenges have passed, the combination of higher productivity and reduced cost of in-person office space could be a major benefit to employers.

While we expect a permanent increase in remote work, our data does not indicate that decision making about telecommuting is a fundamentally different process after the pandemic than it was before. The type of person who gets the option to work from home and takes it is more or less the same after COVID-19 as it was before. The main statistically significant changes in the correlates of remote work between the pre-COVID period and post-pandemic expectations are that those in administrative or professional jobs become more likely to have the option to telecommute.

**Social equity implications**

While many workers gained the ability to telecommute during the COVID-19 pandemic, this new opportunity was unevenly extended to certain groups of workers. The demographics of new telecommuters are similar to those of workers who previously telecommuted; in short, this group is overwhelmingly well-educated, high-income, and working in a professional position. While these workers enjoyed both increased work location flexibility and protection from the potential spread of COVID-19 in the workplace, less educated, lower-income, non-professional workers were more likely to be exposed and did not benefit from increased work location flexibility.

The question arises of how work location flexibility is projected to be distributed across gender and race/ethnicity groups in the post-COVID future, and how that differs from its pre-pandemic distribution. We begin by noting that race, ethnicity, and gender are not major predictors of telecommuting ability in any period in our multivariate models. Rather, education, income, and job category dominate. Nevertheless, bivariate analyses are important for understanding equity outcomes.

The story that emerges from the COVID Future Panel Survey is that there are differences in the ability to telecommute both pre- and post-COVID between race and ethnicity groups (specifically, we compared non-Hispanic white respondents to Hispanic and black respondents), as well as between men and women. These differences are relatively small (between 3 and 5 percentage points), however, and are the same magnitude in both periods. In other words, our data suggest that all of these demographic groups had similar percentage point gains in the ability to telecommute.

**Implications for transport and the environment**

Will a long-term increase in telecommuting have long-term effects on the transport sector? We posit that the answer is yes, but that these effects may be difficult to measure, especially using metrics such as vehicle miles traveled and minutes of congestion delay. In Salon et al. (2021b), we used the COVID Future survey data to estimate that expected increases in telecommuting would result in a 15% reduction in weekly commute distance traveled by car and a 20% drop in transit commute trips, relative to pre-pandemic levels. Similar calculations based on Wave 2 and 3 yield predictions of a 10% and 20% reduction in weekly commute distance traveled by car. The Wave 2 and 3 samples are too small to calculate robust comparative results for changes in transit ridership.

These projected reductions in commute travel are substantial, and if nothing else changed, would clearly mean reduced vehicle miles traveled, reduced peak hour traffic congestion, and reduced transit use. We know, however, that other things will change. Namely, prior research has shown that telecommuters will make other trips during their workdays (Shabanpour et al., 2018) and a reduction in peak-hour congestion will induce demand (Downs, 2004) – especially on the roads in congested urban areas.

The net results of these changes may be that traditional road system metrics will not be substantially changed in a post-COVID world. Vehicle miles traveled data has already nearly returned to 2019 levels (Federal Highway Administration, n.d.). Travel times on congested roads are very sensitive to increased or reduced travel demand, however (Maerivoet and De Moor, 2005, p. 7). If telecommuting and flexible work arrangements lead to even a slightly smaller or more spread out peak, large decreases in congestion could result even with a relatively small change in vehicle miles traveled. That said, congestion delay estimates are creeping back to 2019 levels ("TomTom Traffic Index," n.d.), suggesting that an induced demand effect may yet bring congestion back to near pre-pandemic levels.

Transit systems have been much slower to return to pre-COVID ridership levels, and transit agencies are considering how to restructure their routes and schedules into the future to better serve their riders (Snyder, 2021).

Those who make a long-term switch to telecommuting every day may decide to move farther from their former workplaces since they no longer need to commute regularly. This could be another important transportation outcome of the pandemic-induced telecommuting increase. This is one result from the COVID Future Panel Survey that has changed substantially with each survey wave. Wave 1 data suggested that this would be only a small minority of the US workforce (3 percentage points more than pre-COVID). In Waves 2 and 3, however, those who expect to be everyday telecommuters rose to 5 and then 9 percentage points higher than the pre-COVID period. These are significant increases, and could lead to substantial changes in home location choices and associated traffic patterns.

Finally, it is important to emphasize that approximately half of U.S. workers cannot telecommute, either because it is impossible to do their jobs remotely or because their employers will not allow them to. The COVID Future survey results suggest that only about half of workers telecommuted even at the height of many pandemic-era travel restrictions and office closures. This reality means that physical commuting is certainly here to stay and that there are limits to the contribution that telecommuting can make to reducing our societal need for transportation.

## ACKNOWLEDGMENTS

This research was supported in part by the National Science Foundation (NSF) RAPID program under grants no. 2030156 and 2029962 and by the Center for Teaching Old Models New Tricks (TOMNET), a University Transportation Center sponsored by the U.S. Department of Transportation through grant no. 69A3551747116, as well as by the Knowledge Exchange for Resilience at Arizona State University. This COVID-19 Working Group effort was also supported by the NSF-funded Social Science Extreme Events Research (SSEER) network and the CONVERGE facility at the Natural Hazards Center at the University of Colorado Boulder (NSF Award #1841338) and the NSF CAREER award under grant no. 155173. Any opinions, findings, and conclusions or recommendations expressed in this material are those of the authors and do not necessarily reflect the views of the funders.

**COMPETING INTEREST**

The authors declare that they have no financial or non-financial competing interests.

**AUTHOR CONTRIBUTIONS**

The authors confirm contribution to the paper as follows: study conception and design: D. Salon, M. Bhagat-Conway, R. Pendyala, S. Derrible, A. Mohammadian, E. Rahimi, D. da Silva Baker; data collection: D. Salon, M. Bhagat-Conway, R. Chauhan, D. da Silva Baker; analysis and interpretation of results: D. Salon, M. Bhagat-Conway, L. Mirtich, E. Rahimi; draft manuscript preparation: D. Salon, M. Bhagat-Conway, L. Mirtich, A. Costello. All authors reviewed the results and approved the final version of the manuscript.

**APPENDIX**

**Table A-1: Summary Statistics for Telecommuting Outcomes and Non-Weighting Variables in Three Survey Waves**

| | Wave 1 (July – October 2020) | Wave 2 (November 2020 - March 2021) | Wave 3 (November 2021) |
|---|---|---|---|
| *Pre-Pandemic* | | | |
| Able to Telecommute | 34% | 34% | 34% |
| *Telecommuting Frequency* | | | |
| Never/Rarely Telecommute | 77% | 77% | 77% |
| Sometimes Telecommute | 16% | 16% | 16% |
| Telecommute Every Day | 7% | 7% | 7% |
| *During Pandemic* | | | |
| Able to Telecommute | 54% | 49% | 52% |
| *Telecommuting Frequency* | | | |
| Never/Rarely Telecommute | 49% | 53% | 61% |
| Sometimes Telecommute | 17% | 14% | 19% |
| Telecommute Every Day | 35% | 33% | 20% |
| *Post-Pandemic* | | | |
| Able to Telecommute: Post | 45% | 43% | 53% |
| *Telecommuting Frequency: Post* | | | |
| Never/Rarely Telecommute | 59% | 59% | 51% |
| Sometimes Telecommute | 30% | 29% | 32% |
| Telecommute Every Day | 10% | 12% | 17% |
| *Job category* | | | |
| Other | 30% | 24% | 27% |
| Frontline | 34% | 34% | 28% |
| Professional | 22% | 26% | 24% |
| Education | 9% | 10% | 14% |
| Administrative | 5% | 6% | 7% |
| Student | 19% | 17% | 15% |
| Dense urban resident | 13% | 10% | 13% |
| Full time worker | 77% | 72% | 69% |
| Job change | 9% | 12% | 13% |
| *Attitudes* | | | |
| Enjoy Workplace Interaction | 68% | 70% | 66% |
| Hard to Motivate | 31% | 29% | 24% |
| Enjoy Telecommuting | 53% | 54% | 66% |
| High Speed Internet at Home | 95% | 95% | 95% |
| Extra Bedroom at Home | 59% | 61% | 59% |
| Walk/Bike commuter pre-pandemic | 4% | 4% | 2% |
| Transit commuter pre-pandemic | 8% | 7% | 7% |
| | | | |
| Observations | 4,895 | 1,409 | 1,205 |

**Table A-2: Weighted Post-COVID Telecommuting Expectations by Telecommuting Experience Categories, Wave 2**

| | Telecommuted Pre-COVID | New Telecommuter | | Never Telecommuted |
| | | Chose Not to Telecommute Pre-COVID | Unable to Telecommute Pre-COVID | |
| **Wave 2 Post-COVID Expectation** | | | | |
|---|---|---|---|---|
| Unable to Telecommute | 13% | 19% | 38% | 90% |
| Choose Not to Telecommute | 1% | 4% | 2% | 2% |
| Telecommute Sometimes | 47% | 66% | 48% | 8% |
| Telecommute Every Day | 39% | 11% | 12% | 0% |
| Telecommute if Able | 99% | 95% | 94% | 85% |
| Weighted % of all Workers | 22% | 8% | 19% | 51% |
| N (unweighted) | 399 | 111 | 318 | 582 |

**Table A-3: Weighted Post-COVID Telecommuting Expectations by Telecommuting Experience Categories, Wave 3**

| | Telecommuted Pre-COVID | New Telecommuter | | Never Telecommuted |
| | | Chose Not to Telecommute Pre-COVID | Unable to Telecommute Pre-COVID | |
| **Wave 3 Post-COVID Expectation** | | | | |
|---|---|---|---|---|
| Unable to Telecommute | 9% | 8% | 24% | 83% |
| Choose Not to Telecommute | 1% | 5% | 2% | 5% |
| Telecommute Sometimes | 46% | 71% | 52% | 11% |
| Telecommute Every Day | 44% | 17% | 23% | 1% |
| Telecommute if Able | 99% | 94% | 97% | 68% |
| Weighted % of all Workers | 23% | 8% | 21% | 48% |
| N (unweighted) | 333 | 104 | 266 | 507 |

Notes: The Wave 2 and Wave 3 models of the pre-pandemic period are based on data from the Wave 1 survey, but estimated on the subsamples from Wave 1 that also participated in Wave 2 or 3. We did not ask questions about pre-pandemic choices in our Wave 2 or Wave 3 surveys.

Both the pre-pandemic model for the Wave 2 data and the post-pandemic model for the Wave 3 did not converge when we attempted to estimate telecommuting ability and frequency in a joint model. The marginal effects listed in the tables below, then, are the results of a binary probit model for telecommuting ability and a separate multinomial probit model for telecommuting frequency. This is why the sample size for the telecommuting frequency model is so much smaller for the pre-pandemic period in Wave 2 and the post-pandemic period for Wave 3; it includes only those workers who had the option to telecommute in that period.

**Table A-4: Weighted Marginal Effects on the Option to Telecommute in Three Periods, Wave 2**

| | Pre-pandemic | S.E. | During pandemic | S.E. | Expected Post-pandemic | S.E. |
|---|---|---|---|---|---|---|
| *Educational attainment* | | | | | | |
| Less than bachelors | base | | base | | base | |
| Bachelors | 0.06** | 0.03 | 0.12*** | 0.03 | 0.10*** | 0.03 |
| Graduate | 0.17*** | 0.04 | 0.26*** | 0.04 | 0.20*** | 0.04 |
| *Job category* | | | | | | |
| Other | base | | base | | base | |
| Frontline | -0.21*** | 0.03 | -0.24*** | 0.04 | -0.17*** | 0.04 |
| Professional | 0.21*** | 0.04 | 0.25*** | 0.04 | 0.24*** | 0.04 |
| Education | -0.13*** | 0.04 | NS | | -0.13*** | 0.04 |
| Administrative | NS | | NS | | NS | |
| *Household income* | | | | | | |
| Under $50,000 | base | | base | | base | |
| $50-$100,000 | NS | | NS | | NS | |
| Over $100,000 | 0.08* | 0.04 | 0.10** | 0.04 | 0.09* | 0.05 |
| *Race/Ethnicity* | | | | | | |
| White/Other | base | | base | | base | |
| Hispanic | NS | | NS | | NS | |
| Black | NS | | NS | | NS | |
| Asian | NS | | NS | | NS | |
| *Age category* | | | | | | |
| 18-24 | base | | base | | base | |
| 25-34 | NS | | NS | | NS | |
| 35-49 | NS | | NS | | NS | |
| 50-64 | NS | | NS | | NS | |
| 65+ | NS | | NS | | NS | |
| Student | 0.10** | 0.04 | 0.10** | 0.04 | 0.16*** | 0.04 |
| Dense urban resident | 0.11*** | 0.04 | 0.08* | 0.05 | NS | |
| Full time worker | NS | | 0.06** | 0.03 | -0.05* | 0.03 |
| Job change | NA | | NS | | 0.08* | 0.04 |
| *Region* | | | | | | |
| Arizona | base | | base | | base | |
| New England | NS | | NS | | NS | |
| Middle Atlantic | NS | | NS | | NS | |
| South Atlantic | -0.08* | 0.05 | NS | | NS | |
| East North Central | NS | | NS | | NS | |
| West and East South Central | -0.11** | 0.05 | NS | | NS | |
| West North Central | NS | | NS | | NS | |
| Mountain (no AZ) | NS | | NS | | NS | |
| Pacific | NS | | NS | | NS | |
| Observations | 1,343 | | 1,219 | | 1,387 | |

NS = "Not Significant," NA = "Not Applicable"
*** p<0.01, ** p<0.05, * p<0.1

**Table A-5: Weighted Marginal Effects on Telecommuting Frequency in Three Periods, Wave 2**

| | Pre-pandemic | S.E. | During pandemic | S.E. | Expected Post-pandemic | S.E. |
|---|---|---|---|---|---|---|
| Student | | | | | | |
|    Never/Rarely Telecommute | -0.06 | 0.06 | -0.03 | 0.04 | NS | |
|    Sometimes Telecommute | 0.15** | 0.07 | 0.15*** | 0.06 | NS | |
|    Telecommute Every Day | -0.10 | 0.06 | -0.12** | 0.06 | NS | |
| *Job Category* | | | | | | |
| Other | base | | base | | base | |
| Frontline | | | | | | |
|    Never/Rarely Telecommute | NS | | NS | | -0.02 | 0.04 |
|    Sometimes Telecommute | NS | | NS | | 0.09** | 0.04 |
|    Telecommute Every Day | NS | | NS | | -0.06** | 0.03 |
| Professional | | | | | | |
|    Never/Rarely Telecommute | 0.10** | 0.05 | NS | | NS | |
|    Sometimes Telecommute | 0.05 | 0.06 | NS | | NS | |
|    Telecommute Every Day | -0.15*** | 0.05 | NS | | NS | |
| Education | | | | | | |
|    Never/Rarely Telecommute | 0.03 | 0.06 | NS | | -0.01 | 0.04 |
|    Sometimes Telecommute | 0.18** | 0.07 | NS | | 0.12** | 0.05 |
|    Telecommute Every Day | -0.22*** | 0.05 | NS | | -0.11*** | 0.04 |
| Administrative | | | | | | |
|    Never/Rarely Telecommute | NS | | -0.03 | 0.03 | -0.11 | 0.07 |
|    Sometimes Telecommute | NS | | 0.13* | 0.08 | 0.18*** | 0.07 |
|    Telecommute Every Day | NS | | -0.10 | 0.08 | -0.07* | 0.04 |
| *Attitudes* | | | | | | |
| Enjoy Workplace Interaction | | | | | | |
|    Never/Rarely Telecommute | 0.05*** | 0.02 | NS | | 0.00 | 0.01 |
|    Sometimes Telecommute | -0.03 | 0.02 | NS | | 0.02 | 0.02 |
|    Telecommute Every Day | -0.03 | 0.02 | NS | | -0.02** | 0.01 |
| Hard to Motivate | | | | | | |
|    Never/Rarely Telecommute | NS | | NS | | -0.02 | 0.01 |
|    Sometimes Telecommute | NS | | NS | | 0.03* | 0.02 |
|    Telecommute Every Day | NS | | NS | | -0.01 | 0.01 |
| Enjoy Telecommuting | | | | | | |
|    Never/Rarely Telecommute | -0.10*** | 0.02 | -0.02 | 0.02 | -0.08** | 0.04 |
|    Sometimes Telecommute | -0.04 | 0.03 | -0.09*** | 0.02 | 0.00 | 0.05 |
|    Telecommute Every Day | 0.14*** | 0.03 | 0.11*** | 0.02 | 0.07*** | 0.02 |
| Dense Urban Resident | | | | | | |
|    Never/Rarely Telecommute | NS | | NS | | NS | |
|    Sometimes Telecommute | NS | | NS | | NS | |
|    Telecommute Every Day | NS | | NS | | NS | |

|  | Pre-pandemic | S.E. | During pandemic | S.E. | Expected Post-pandemic | S.E. |
|---|---|---|---|---|---|---|
| High Speed Internet at Home |  |  |  |  |  |  |
|   Never/Rarely Telecommute | NS |  | NS |  | NS |  |
|   Sometimes Telecommute | NS |  | NS |  | NS |  |
|   Telecommute Every Day | NS |  | NS |  | NS |  |
| Extra Bedroom at Home |  |  |  |  |  |  |
|   Never/Rarely Telecommute | NS |  | NS |  | 0.02 | 0.03 |
|   Sometimes Telecommute | NS |  | NS |  | 0.03 | 0.04 |
|   Telecommute Every Day | NS |  | NS |  | -0.05* | 0.02 |
| Female |  |  |  |  |  |  |
|   Never/Rarely Telecommute | NS |  | NS |  | NS |  |
|   Sometimes Telecommute | NS |  | NS |  | NS |  |
|   Telecommute Every Day | NS |  | NS |  | NS |  |
| Children |  |  |  |  |  |  |
|   Never/Rarely Telecommute | NS |  | NS |  | 0.02 | 0.03 |
|   Sometimes Telecommute | NS |  | NS |  | 0.02 | 0.04 |
|   Telecommute Every Day | NS |  | NS |  | -0.04* | 0.02 |
| *Race/Ethnicity* |  |  |  |  |  |  |
| White/Other | base |  | base |  | base |  |
| Hispanic |  |  |  |  |  |  |
|   Never/Rarely Telecommute | NS |  | NS |  | NS |  |
|   Sometimes Telecommute | NS |  | NS |  | NS |  |
|   Telecommute Every Day | NS |  | NS |  | NS |  |
| Black |  |  |  |  |  |  |
|   Never/Rarely Telecommute | NS |  | NS |  | NS |  |
|   Sometimes Telecommute | NS |  | NS |  | NS |  |
|   Telecommute Every Day | NS |  | NS |  | NS |  |
| Asian |  |  |  |  |  |  |
|   Never/Rarely Telecommute | 0.10 | 0.07 | -0.00 | 0.03 | 0.09** | 0.05 |
|   Sometimes Telecommute | 0.06 | 0.08 | -0.16** | 0.07 | -0.00 | 0.07 |
|   Telecommute Every Day | -0.16** | 0.07 | 0.16** | 0.07 | -0.09** | 0.04 |
| *Age category* |  |  |  |  |  |  |
| 18-24 | base |  | base |  | base |  |
| 25-34 |  |  |  |  |  |  |
|   Never/Rarely Telecommute | NS |  | NS |  | 0.02 | 0.05 |
|   Sometimes Telecommute | NS |  | NS |  | -0.10* | 0.06 |
|   Telecommute Every Day | NS |  | NS |  | 0.08** | 0.03 |
| 35-49 |  |  |  |  |  |  |
|   Never/Rarely Telecommute | NS |  | NS |  | 0.05 | 0.05 |
|   Sometimes Telecommute | NS |  | NS |  | -0.16*** | 0.06 |
|   Telecommute Every Day | NS |  | NS |  | 0.11*** | 0.03 |
| 50-64 |  |  |  |  |  |  |
|   Never/Rarely Telecommute | NS |  | -0.09 | 0.11 | 0.10* | 0.06 |
|   Sometimes Telecommute | NS |  | -0.11 | 0.12 | -0.19*** | 0.07 |

| | Pre-pandemic | S.E. | During pandemic | S.E. | Expected Post-pandemic | S.E. |
|---|---|---|---|---|---|---|
| Telecommute Every Day | NS | | 0.20* | 0.11 | 0.09*** | 0.04 |
| 65+ | | | | | | |
|   Never/Rarely Telecommute | NS | | NS | | 0.08 | 0.07 |
|   Sometimes Telecommute | NS | | NS | | -0.20*** | 0.07 |
|   Telecommute Every Day | NS | | NS | | 0.13*** | 0.05 |
| *Job/Commute Characteristics* | | | | | | |
| Nonmotorized commuter pre-pandemic | | | | | | |
|   Never/Rarely Telecommute | NS | | NS | | NS | |
|   Sometimes Telecommute | NS | | NS | | NS | |
|   Telecommute Every Day | NS | | NS | | NS | |
| Transit commuter pre-pandemic | | | | | | |
|   Never/Rarely Telecommute | -0.01 | 0.06 | NS | | -0.09 | 0.07 |
|   Sometimes Telecommute | 0.28*** | 0.08 | NS | | 0.19*** | 0.07 |
|   Telecommute Every Day | -0.27*** | 0.08 | NS | | -0.10** | 0.04 |
| Full time | | | | | | |
|   Never/Rarely Telecommute | 0.11** | 0.05 | -0.00 | 0.02 | -0.04 | 0.03 |
|   Sometimes Telecommute | -0.02 | 0.06 | -0.11** | 0.05 | 0.08** | 0.03 |
|   Telecommute Every Day | -0.10** | 0.04 | 0.12** | 0.05 | -0.04* | 0.02 |
| Job Change | | | | | | |
|   Never/Rarely Telecommute | NA | | NS | | NS | |
|   Sometimes Telecommute | NA | | NS | | NS | |
|   Telecommute Every Day | NA | | NS | | NS | |
| *Region* | | | | | | |
| Arizona | base | | base | | base | |
| New England | | | | | | |
|   Never/Rarely Telecommute | NS | | NS | | 0.07 | 0.07 |
|   Sometimes Telecommute | NS | | NS | | 0.06 | 0.09 |
|   Telecommute Every Day | NS | | NS | | -0.13** | 0.06 |
| Middle Atlantic | | | | | | |
|   Never/Rarely Telecommute | -0.01 | 0.08 | NS | | 0.02 | 0.06 |
|   Sometimes Telecommute | 0.22** | 0.10 | NS | | 0.07 | 0.08 |
|   Telecommute Every Day | -0.21*** | 0.08 | NS | | -0.09* | 0.05 |
| South Atlantic | | | | | | |
|   Never/Rarely Telecommute | -0.07 | 0.07 | NS | | -0.05 | 0.05 |
|   Sometimes Telecommute | 0.23*** | 0.08 | NS | | 0.12* | 0.06 |
|   Telecommute Every Day | -0.16** | 0.06 | NS | | -0.07 | 0.04 |
| East North Central | | | | | | |
|   Never/Rarely Telecommute | 0.05 | 0.07 | NS | | -0.05 | 0.05 |
|   Sometimes Telecommute | 0.10 | 0.08 | NS | | 0.14** | 0.06 |
|   Telecommute Every Day | -0.15** | 0.07 | NS | | -0.09** | 0.04 |
| West and East South | | | | | | |

| | Pre-pandemic | S.E. | During pandemic | S.E. | Expected Post-pandemic | S.E. |
|---|---|---|---|---|---|---|
| Central | | | | | | |
| Never/Rarely Telecommute | 0.10 | 0.09 | NS | | -0.01 | 0.06 |
| Sometimes Telecommute | 0.07 | 0.10 | NS | | 0.15* | 0.08 |
| Telecommute Every Day | -0.17** | 0.08 | NS | | -0.14*** | 0.05 |
| West North Central | | | | | | |
| Never/Rarely Telecommute | NS | | NS | | -0.13* | 0.08 |
| Sometimes Telecommute | NS | | NS | | 0.18** | 0.08 |
| Telecommute Every Day | NS | | NS | | -0.05 | 0.05 |
| Mountain (no AZ) | | | | | | |
| Never/Rarely Telecommute | NS | | NS | | NS | |
| Sometimes Telecommute | NS | | NS | | NS | |
| Telecommute Every Day | NS | | NS | | NS | |
| Pacific | | | | | | |
| Never/Rarely Telecommute | -0.04 | 0.07 | NS | | NS | |
| Sometimes Telecommute | 0.20** | 0.08 | NS | | NS | |
| Telecommute Every Day | -0.16** | 0.07 | NS | | NS | |
| | | | | | | |
| Observations | 540 | | 1,219 | | 1,387 | |

NS = "Not Significant," NA = "Not Applicable"

*** p<0.01, ** p<0.05, * p<0.1

**Table A-6: Weighted Marginal Effects on the Option to Telecommute in Three Periods, Wave 3**

| | Pre-pandemic | S.E. | During pandemic (Fall 2021) | S.E. | Expected Post-pandemic | S.E. |
|---|---|---|---|---|---|---|
| *Educational attainment* | | | | | | |
| Less than bachelors | base | | base | | base | |
| Bachelors | 0.09*** | 0.03 | 0.13*** | 0.03 | 0.08** | 0.03 |
| Graduate | 0.16*** | 0.04 | 0.20*** | 0.04 | 0.16*** | 0.04 |
| *Job category* | | | | | | |
| Other | base | | base | | base | |
| Frontline | -0.23*** | 0.04 | -0.15*** | 0.04 | -0.16*** | 0.04 |
| Professional | 0.17*** | 0.04 | 0.29*** | 0.04 | 0.28*** | 0.04 |
| Education | -0.15*** | 0.04 | NS | | -0.11** | 0.05 |
| Administrative | NS | | 0.14*** | 0.05 | 0.12** | 0.05 |
| *Household income* | | | | | | |
| Under $50,000 | base | | base | | base | |
| $50-$100,000 | NS | | 0.11*** | 0.04 | NS | |
| Over $100,000 | 0.09** | 0.04 | 0.20*** | 0.04 | 0.15*** | 0.04 |
| *Race/Ethnicity* | | | | | | |
| White/Other | base | | base | | base | |
| Hispanic | NS | | NS | | NS | |
| Black | NS | | NS | | NS | |
| Asian | NS | | NS | | NS | |
| *Age category* | | | | | | |
| 18-24 | base | | base | | base | |
| 25-34 | NS | | NS | | NS | |
| 35-49 | NS | | NS | | NS | |
| 50-64 | NS | | NS | | -0.17** | 0.07 |
| 65+ | NS | | NS | | -0.24*** | 0.07 |
| Student | NS | | NS | | NS | |
| Dense urban resident | 0.09** | 0.04 | 0.10** | 0.04 | NS | |
| Full time worker | NS | | 0.05* | 0.03 | NS | |
| Job change | NA | | NS | | NS | |
| *Region* | | | | | | |
| Arizona | base | | base | | base | |
| New England | NS | | NS | | NS | |
| Middle Atlantic | NS | | NS | | NS | |
| South Atlantic | NS | | NS | | NS | |
| East North Central | -0.11** | 0.05 | NS | | NS | |
| West and East South Central | NS | | NS | | NS | |
| West North Central | NS | | NS | | NS | |
| Mountain (no AZ) | NS | | NS | | NS | |
| Pacific | NS | | NS | | NS | |
| | | | | | | |
| Observations | 1,258 | | 1,196 | | 1,198 | |

NS = "Not Significant," NA = "Not Applicable"
*** p<0.01, ** p<0.05, * p<0.1

**Table A-7: Weighted Marginal Effects on Telecommuting Frequency in Three Periods, Wave 3**

| | Pre-pandemic | S.E. | During pandemic (Fall 2021) | S.E. | Expected Post-pandemic | S.E. |
|---|---|---|---|---|---|---|
| Student | | | | | | |
|   Never/Rarely Telecommute | NS | | NS | | NS | |
|   Sometimes Telecommute | NS | | NS | | NS | |
|   Telecommute Every Day | NS | | NS | | NS | |
| *Job Category* | | | | | | |
| Other | base | | base | | base | |
| Frontline | | | | | | |
|   Never/Rarely Telecommute | -0.06 | 0.07 | 0.03 | 0.04 | 0.04 | 0.03 |
|   Sometimes Telecommute | 0.14* | 0.08 | 0.06 | 0.06 | 0.06 | 0.06 |
|   Telecommute Every Day | -0.09*** | 0.03 | -0.09* | 0.05 | -0.10* | 0.05 |
| Professional | | | | | | |
|   Never/Rarely Telecommute | NS | | 0.04 | 0.04 | NS | |
|   Sometimes Telecommute | NS | | -0.12* | 0.06 | NS | |
|   Telecommute Every Day | NS | | 0.08 | 0.05 | NS | |
| Education | | | | | | |
|   Never/Rarely Telecommute | 0.06 | 0.08 | NS | | 0.00 | 0.03 |
|   Sometimes Telecommute | 0.02 | 0.08 | NS | | 0.15** | 0.06 |
|   Telecommute Every Day | -0.08** | 0.03 | NS | | -0.15*** | 0.05 |
| Administrative | | | | | | |
|   Never/Rarely Telecommute | NS | | NS | | 0.02 | 0.04 |
|   Sometimes Telecommute | NS | | NS | | 0.10 | 0.07 |
|   Telecommute Every Day | NS | | NS | | -0.12** | 0.06 |
| *Attitudes* | | | | | | |
| Enjoy Workplace Interaction | | | | | | |
|   Never/Rarely Telecommute | 0.08*** | 0.02 | 0.02* | 0.01 | -0.01 | 0.01 |
|   Sometimes Telecommute | -0.06*** | 0.02 | 0.01 | 0.02 | 0.11*** | 0.02 |
|   Telecommute Every Day | -0.02* | 0.01 | -0.03** | 0.01 | -0.09*** | 0.01 |
| Hard to Motivate | | | | | | |
|   Never/Rarely Telecommute | 0.00 | 0.02 | 0.02 | 0.01 | NS | |
|   Sometimes Telecommute | 0.02 | 0.02 | 0.03 | 0.02 | NS | |
|   Telecommute Every Day | -0.02* | 0.01 | -0.04*** | 0.02 | NS | |
| Enjoy Telecommuting | | | | | | |
|   Never/Rarely Telecommute | -0.08*** | 0.02 | -0.08*** | 0.02 | -0.07*** | 0.01 |
|   Sometimes Telecommute | 0.05* | 0.03 | 0.04* | 0.02 | -0.05* | 0.02 |
|   Telecommute Every Day | 0.03** | 0.01 | 0.05*** | 0.02 | 0.11*** | 0.02 |
| Dense Urban Resident | | | | | | |
|   Never/Rarely Telecommute | NS | | NS | | -0.06 | 0.04 |
|   Sometimes Telecommute | NS | | NS | | 0.11* | 0.06 |
|   Telecommute Every Day | NS | | NS | | -0.06 | 0.05 |

| | Pre-pandemic | S.E. | During pandemic (Fall 2021) | S.E. | Expected Post-pandemic | S.E. |
|---|---|---|---|---|---|---|
| High Speed Internet at Home | | | | | | |
|   Never/Rarely Telecommute | NS | | NS | | NS | |
|   Sometimes Telecommute | NS | | NS | | NS | |
|   Telecommute Every Day | NS | | NS | | NS | |
| Extra Bedroom at Home | | | | | | |
|   Never/Rarely Telecommute | NS | | NS | | NS | |
|   Sometimes Telecommute | NS | | NS | | NS | |
|   Telecommute Every Day | NS | | NS | | NS | |
| Female | | | | | | |
|   Never/Rarely Telecommute | NS | | NS | | NS | |
|   Sometimes Telecommute | NS | | NS | | NS | |
|   Telecommute Every Day | NS | | NS | | NS | |
| Children | | | | | | |
|   Never/Rarely Telecommute | NS | | 0.06* | 0.03 | NS | |
|   Sometimes Telecommute | NS | | 0.00 | 0.04 | NS | |
|   Telecommute Every Day | NS | | -0.06* | 0.03 | NS | |
| *Race/Ethnicity* | | | | | | |
| White/Other | base | | base | | base | |
| Hispanic | | | | | | |
|   Never/Rarely Telecommute | NS | | 0.09** | 0.05 | NS | |
|   Sometimes Telecommute | NS | | 0.05 | 0.06 | NS | |
|   Telecommute Every Day | NS | | -0.14*** | 0.05 | NS | |
| Black | | | | | | |
|   Never/Rarely Telecommute | NS | | 0.09* | 0.05 | NS | |
|   Sometimes Telecommute | NS | | -0.04 | 0.07 | NS | |
|   Telecommute Every Day | NS | | -0.05 | 0.05 | NS | |
| Asian | | | | | | |
|   Never/Rarely Telecommute | NS | | NS | | -0.13*** | 0.05 |
|   Sometimes Telecommute | NS | | NS | | 0.18*** | 0.07 |
|   Telecommute Every Day | NS | | NS | | -0.05 | 0.06 |
| *Age category* | | | | | | |
| 18-24 | base | | base | | base | |
| 25-34 | | | | | | |
|   Never/Rarely Telecommute | 0.29** | 0.13 | 0.15* | 0.08 | NS | |
|   Sometimes Telecommute | -0.19 | 0.14 | 0.05 | 0.12 | NS | |
|   Telecommute Every Day | -0.10 | 0.08 | -0.20* | 0.11 | NS | |
| 35-49 | | | | | | |
|   Never/Rarely Telecommute | NS | | NS | | -0.01 | 0.04 |
|   Sometimes Telecommute | NS | | NS | | -0.11 | 0.08 |
|   Telecommute Every Day | NS | | NS | | 0.12* | 0.07 |
| 50-64 | | | | | | |
|   Never/Rarely Telecommute | 0.20* | 0.12 | 0.11 | 0.07 | 0.00 | 0.04 |
|   Sometimes Telecommute | -0.15 | 0.13 | 0.10 | 0.11 | -0.17** | 0.08 |

| | Pre-pandemic | S.E. | During pandemic (Fall 2021) | S.E. | Expected Post-pandemic | S.E. |
|---|---|---|---|---|---|---|
| Telecommute Every Day | -0.06 | 0.08 | -0.20* | 0.11 | 0.16** | 0.07 |
| 65+ | | | | | | |
|   Never/Rarely Telecommute | NS | | 0.16* | 0.08 | -0.03 | 0.05 |
|   Sometimes Telecommute | NS | | 0.12 | 0.12 | -0.17* | 0.09 |
|   Telecommute Every Day | NS | | -0.28** | 0.11 | 0.21** | 0.08 |
| *Job/Commute Characteristics* | | | | | | |
| Nonmotorized commuter pre-pandemic | | | | | | |
|   Never/Rarely Telecommute | NS | | NS | | NS | |
|   Sometimes Telecommute | NS | | NS | | NS | |
|   Telecommute Every Day | NS | | NS | | NS | |
| Transit commuter pre-pandemic | | | | | | |
|   Never/Rarely Telecommute | -0.03 | 0.07 | -0.10* | 0.05 | -0.14** | 0.07 |
|   Sometimes Telecommute | 0.13 | 0.08 | 0.08 | 0.06 | 0.21*** | 0.07 |
|   Telecommute Every Day | -0.10** | 0.05 | 0.02 | 0.05 | -0.07 | 0.06 |
| Full time | | | | | | |
|   Never/Rarely Telecommute | NS | | -0.00 | 0.04 | -0.04 | 0.02 |
|   Sometimes Telecommute | NS | | -0.11** | 0.05 | 0.12*** | 0.04 |
|   Telecommute Every Day | NS | | 0.12*** | 0.04 | -0.09** | 0.04 |
| Job Change | | | | | | |
|   Never/Rarely Telecommute | NA | | NS | | -0.03 | 0.03 |
|   Sometimes Telecommute | NA | | NS | | -0.12*** | 0.05 |
|   Telecommute Every Day | NA | | NS | | 0.15*** | 0.04 |
| *Region* | | | | | | |
| Arizona | base | | base | | base | |
| New England | | | | | | |
|   Never/Rarely Telecommute | NS | | NS | | NS | |
|   Sometimes Telecommute | NS | | NS | | NS | |
|   Telecommute Every Day | NS | | NS | | NS | |
| Middle Atlantic | | | | | | |
|   Never/Rarely Telecommute | NS | | NS | | NS | |
|   Sometimes Telecommute | NS | | NS | | NS | |
|   Telecommute Every Day | NS | | NS | | NS | |
| South Atlantic | | | | | | |
|   Never/Rarely Telecommute | NS | | NS | | NS | |
|   Sometimes Telecommute | NS | | NS | | NS | |
|   Telecommute Every Day | NS | | NS | | NS | |
| East North Central | | | | | | |
|   Never/Rarely Telecommute | NS | | NS | | NS | |
|   Sometimes Telecommute | NS | | NS | | NS | |
|   Telecommute Every Day | NS | | NS | | NS | |
| West and East South | | | | | | |

| | Pre-pandemic | S.E. | During pandemic (Fall 2021) | S.E. | Expected Post-pandemic | S.E. |
|---|---|---|---|---|---|---|
| Central | | | | | | |
|   Never/Rarely Telecommute | NS | | NS | | NS | |
|   Sometimes Telecommute | NS | | NS | | NS | |
|   Telecommute Every Day | NS | | NS | | NS | |
| West North Central | | | | | | |
|   Never/Rarely Telecommute | NS | | -0.11* | 0.06 | NS | |
|   Sometimes Telecommute | NS | | 0.14* | 0.08 | NS | |
|   Telecommute Every Day | NS | | -0.03 | 0.07 | NS | |
| Mountain (no AZ) | | | | | | |
|   Never/Rarely Telecommute | NS | | NS | | NS | |
|   Sometimes Telecommute | NS | | NS | | NS | |
|   Telecommute Every Day | NS | | NS | | NS | |
| Pacific | | | | | | |
|   Never/Rarely Telecommute | -0.14* | 0.08 | NS | | NS | |
|   Sometimes Telecommute | 0.16** | 0.08 | NS | | NS | |
|   Telecommute Every Day | -0.02 | 0.03 | NS | | NS | |
| | | | | | | |
| Observations | 1,258 | | 1,196 | | 688 | |

NS = "Not Significant," NA = "Not Applicable"

*** p<0.01, ** p<0.05, * p<0.1

**Supplemental Material: Mathematical justification for a sample-selection model**

*Accompanying:* The Effects of the COVID-19 Pandemic on Telecommuting in the United States

In this research, we are interested in understanding the determinants of both telecommuting ability and frequency. To understand the latter choice, we would like to estimate a model of the desired frequency of telecommuting for all workers. We only observe telecommuting frequency for those who have the option to telecommute, however. This creates a sample-selection problem; we can only estimate our model of telecommuting frequency on a non-random subset of the population.

If there are unobserved variables that predict both telecommuting frequency and ability, results of the frequency model will be biased (Wooldridge, 2010). To see why, consider two hypothetical models that include the full population: first, a model of telecommuting ability (which we'll call the selection model, since it selects the sample for the frequency model), and second a model of telecommuting frequency (which we'll call the outcome model, since it predicts an outcome based on the sample selected by the first model). A shared unobserved predictor will result in a correlation between the error terms of the two models, because that predictor will be included in both error terms. The error terms of the two models can be conceptualized as a joint normal distribution, as shown in Figure SM1a, with each margin representing the error term of one of the models. Since the error terms are correlated, those in the selected sample will have above average error terms in the outcome model (assuming a positive correlation).

To see why, consider Figure SM1b. For the sake of simplicity, assume that both the selection and outcome models are binary probit. In the probit model of ability to telecommute, workers with the ability to telecommute have $XB + \epsilon \geq 0$, where X is a vector of observed variables, B is a vector of estimated parameters, and $\epsilon$ is an error term distributed according to one margin of the joint error term in Figure SM1a.

All of the selected workers have $XB + \epsilon \geq 0$, or equivalently $\epsilon \geq -XB$. This implies that the estimation sample for the outcome model will have above-average error terms in the selection model. Since the error terms are correlated, this implies that the error terms for these individuals in the population-level outcome model are also above average—and how much above average depends on the individual. Figure SM1b shows the joint error term for an individual who does have the option to telecommute, and has $XB = -1$. Because the error term in the selection model must be greater than 1, the joint error distribution is truncated. The marginal distribution of this truncated normal distribution along the non-truncated axis has a mean greater than 0 (corresponding with the outcome model error term). Since the distribution will be truncated at different points for different respondents, the mean of the error term in the outcome model will be different for different respondents. This mean error term can be considered an omitted variable (Heckman, 1979), causing bias in the coefficients for the constant and any other variables correlated with the omitted variable.

The most well-known method for accounting for selection bias is Heckman's sample selection model (Heckman, 1979). Heckman uses a probit model of whether an observation is selected (in our case, whether a respondent has the ability to telecommute) to derive information about unobserved common predictors of selection and the outcome of interest. While the unobserved variables that contribute to both the ability to telecommute and telecommuting frequency are by definition unobserved, they contribute to the error term of the probit model of selection. Heckman's model adds a term derived from this error term to the model of the outcome.

Heckman's model is a computationally-tractable method of estimating a more general class of selection models with common unobserved predictors. Importantly, Heckman's model predicts a continuous outcome, while we wish to predict a discrete outcome. The broader problem that Heckman's model solves for the linear case is that of two population-level models with correlated error terms, where the data available to estimate one of the models is predicated on the dependent variable of the other. However, since our measure of telecommuting frequency is not continuous, we cannot use Heckman's method.

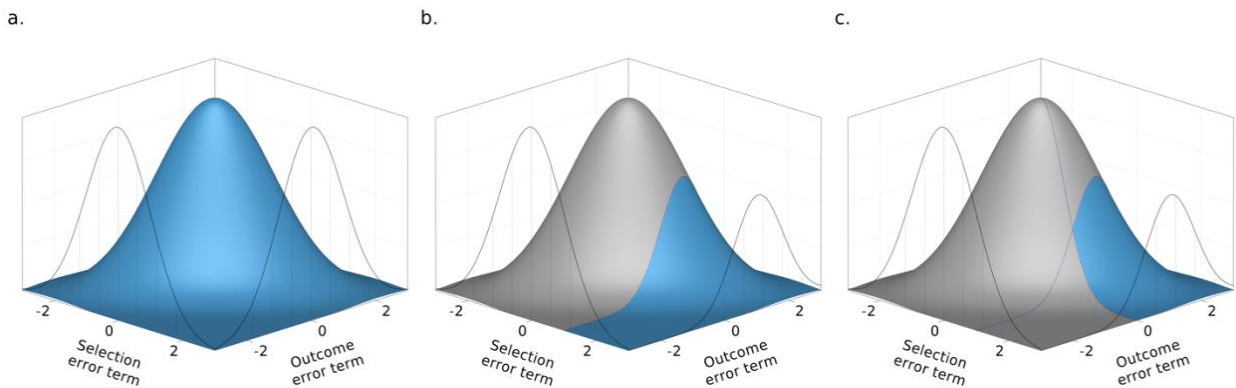

Figure SM1: The multidimensional error term of a probit model with selection.

Early explorations of sample selection models for discrete variables modify Heckman's correction to account for non-unit variance, but do not account for the non-normal distribution, considering their results to be approximations—likely close approximations (Van de Ven & Van

Praag, 1981). However, due to advances in computational power since these early explorations, we can now jointly estimate two probit models with correlated error terms, directly accounting for both the nonzero mean as well as the non-normal distribution and non-unit variance distribution of the correlated error terms.

The model is estimated using maximum likelihood. For a non-selected respondent (i.e. someone who does not have the ability to telecommute), the likelihood is the portion of the joint error term associated with non-selection, i.e. the non-shaded area in Figure SM1a. For a respondent who both has the option to telecommute and telecommutes frequently, the likelihood is the shaded area in Figure SM1c, which corresponds to the joint probability of having the option to telecommute and doing so frequently. For a respondent who has the option to telecommute but does not do so frequently, the likelihood is the remaining area. In the latter two cases, the likelihood is integrated over the truncated joint error term, implicitly accounting for the non-zero mean of the error term relative to a model of the full population, and thus producing consistent population-level estimates. Both the Heckman model and the model we use requires an assumption that the choice process regarding telecommuting frequency would be the same for the people who do not currently have the option to telecommute as it is for those who do currently have that option.

The description above is based on a binary probit for both the models of telecommuting ability and frequency. The model we implement uses a multinomial probit model of telecommuting frequency, dividing telecommuting frequency into three categories. Thus, there are three error terms: one for the model of telecommuting ability, and two for the model of telecommuting frequency (one utility function in the multinomial probit model is held fixed at 0). We assume the latter two error terms are uncorrelated (this is the standard independence of irrelevant alternatives assumption common in discrete choice modeling), while we estimate the correlations between the error term from the ability model and each of the error terms from the frequency model.

The Heckman model requires an exclusion restriction for robust identification of parameters in the outcome and to prevent the model from being identified only by functional form (Wooldridge, 2010). The model we use is no different. In both the Heckman model and our model, the per-observation mean of the error term in the outcome model is a function of the observed covariates in the selection model and the correlation between the error terms of the two models.

For instance, in the example in Figure SM1b, if the predicted value in the selection model was smaller, the joint normal distribution would be truncated at a higher value, and the mean of the error term in the outcome model would be larger. The sensitivity of the error term of the outcome model to the covariates in the selection model is determined by the correlation between the error terms, which is simultaneously estimated with the two models.

Since the mean of the error term in the outcome model is a function of the predicted value and thus the covariates in the selection model, there are collinearity concerns if there are no predictors in the selection model that do not appear in the outcome model. Since it is a nonlinear function, the model may still be identified even without an exclusion restriction, but such identification is only due to the assumed joint normal functional form of the error terms, and may not be robust to model misspecification (Wooldridge, 2010). In our model, we include education and household income in the ability model, and we do not believe these variables affect telecommute frequency conditional on ability as they proxy for job type.

We estimate our model using the -cmp- package in Stata (Roodman, 2011), a package for fitting a variety of simultaneous-equation models including this type of sample selection model.